\documentclass[preprint,12pt]{elsarticle}

\usepackage[utf8]{inputenc}
\usepackage{microtype}
\usepackage{xurl}
\providecommand{\href}[2]{#2}
\usepackage{amsmath}
\usepackage{amssymb}
\usepackage{booktabs}
\usepackage{tabularx}
\usepackage{float}
\usepackage{graphicx}

\journal{Journal of Manufacturing Systems}

\begin{document}

\begin{frontmatter}

\title{Optimistic Verifiable Claims: A Blockchain Protocol for Conditionally Confidential Bidding in Decentralized Manufacturing}

\author[aff1]{Marko Corn\fnref{fn1}}\ead{marko.corn@fs.uni-lj.si}
\author[aff1]{Nejc Rožman\fnref{fn1}}
\author[aff1]{Primož Podržaj\corref{cor1}}

\cortext[cor1]{Corresponding author}
\fntext[fn1]{These authors contributed equally to this work.}

\address[aff1]{Faculty of Mechanical Engineering, University of Ljubljana, A\v{s}ker\v{c}eva 6, 1000 Ljubljana, Slovenia}

\begin{abstract}
Decentralized manufacturing faces a pre-contractual impasse: a Provider cannot price a service accurately without inspecting the design file, yet the Consumer cannot share that file without exposing intellectual property. We introduce the \textbf{Optimistic Verifiable Claim (OVC)}, a blockchain protocol that lets a Consumer publish a verifiable claim about a concealed design (such as the material it consumes) and a Provider price and bid on it without seeing the design. The claim is committed when the service is posted and stands unless the selected Provider challenges it; a challenge triggers a deterministic on-chain check that exposes any dishonesty, and the design is disclosed only to settle a dispute, never on the honest path. We implement four checks (authorized key access, delivery-channel integrity, syntactic conformance, and declared material consumption) in Solidity and measure them on a real 6.41~MB G-code file, the 3DBenchy, across Ethereum, Arbitrum, and opBNB. Every service incurs the cost of posting the encrypted design, with or without a dispute. For the 3DBenchy, the no-dispute outcome costs \$7{,}207 in up to 9~hours on Ethereum, \$288 in 3~min on Arbitrum, and \$2.87 in 2~min on opBNB, and a fully contested dispute costs \$49{,}660 in up to 57~hours on Ethereum, \$1{,}988 in 19~min on Arbitrum, and \$19.73 in 13~min on opBNB. Costs and times grow with size: for a 50~MB industrial design, an undisputed service reaches \$56{,}173 and up to 3~days on Ethereum against \$22.36 and 16~min on opBNB, and a fully contested dispute reaches \$488{,}440 over up to 18~days on Ethereum against \$195 and 1.6~hours on opBNB. Of the four, the material-consumption check is the costliest, its predicate being the most expensive to evaluate on-chain. OVC makes confidential, claim-based bidding economically feasible on Arbitrum and opBNB, but not on Ethereum at industrial scale.
\end{abstract}


\end{frontmatter}

\section{Introduction}

The transition toward Decentralized Manufacturing (DM) is reshaping global production by virtualizing manufacturing resources and exposing them as network-accessible services \cite{Tao2011}. In this emerging peer-to-peer (P2P) economy, production capacity is traded dynamically, allowing Consumers to outsource high-precision manufacturing tasks to a global pool of heterogeneous Providers without the need for centralized intermediaries \cite{Lowe2019, yu_blockchain-based_2020-1}. While blockchain technology has been widely proposed as the coordination layer for these ecosystems due to its auditability and tamper-evident records \cite{Angrish2018, Bodkhe2020}, a critical bottleneck remains in the pre-contractual negotiation phase.

In high-value additive manufacturing, design files such as G-code encode sensitive intellectual property (IP), including proprietary geometries and process-specific toolpaths \cite{Yampolskiy2014}. During the transition from discovery to a binding contract, a ``deadlock of trust'' occurs, a pre-contractual manifestation of Arrow's information paradox \cite{arrow_economic_1962}: the Consumer requires an accurate bid based on the technical complexity of the task, yet the Provider cannot generate such a bid without inspecting the design file, the very information the Consumer seeks to protect. Current literature typically assumes that task descriptions are either fully disclosed or handled by trusted brokers \cite{Lanza2019, Chen2019}. However, in open, trust-minimized marketplaces, these assumptions are often incompatible with the competitive need for data confidentiality and IP protection.

Existing cryptographic solutions address either confidentiality or delivery integrity, but not the \emph{veracity} of what is delivered. Access-control schemes such as Attribute-Based Encryption restrict who may read a file \cite{Fan2022, zhang_blockchain-enabled_2024} while guaranteeing nothing about the claims made of it; fair-exchange and escrow mechanisms guarantee atomic delivery but remain content-agnostic \cite{asokan_optimistic_2000}. A smart contract acting as arbiter can verify that a bitstring matches a hash, yet cannot determine whether it is syntactically valid G-code or whether it consumes the material its price was based on.

This paper addresses a specific gap in decentralized manufacturing: the absence of a mechanism that makes a claim about a concealed artifact \textbf{binding and provably falsifiable} without a trusted intermediary. We introduce the \textbf{Optimistic Verifiable Claim (OVC)} framework, a modular blockchain primitive for the optimistic adjudication of committed manufacturing claims. OVC binds a claim at commitment time and guarantees that, if challenged, a deterministic on-chain predicate exposes any dishonest claim. The Provider prices from the published claim (for example, declared material consumption), then receives the artifact during Specification Handover and verifies it before manufacturing begins; a detected mismatch is disputed in handover and the workflow terminates before Execution. A false claim is therefore provable whenever it is challenged, and the resulting verdict is attributed to the party at fault and recorded on-chain, where a marketplace can consume it as evidence for penalties or reputation.

The framework provides \textbf{optimistic confidentiality}: a claim is accepted unless challenged, and the committed artifact stays private on the no-dispute path. Adjudication is the deliberate exception: to prove a contested claim, the Consumer discloses the witness on-chain, where a deterministic predicate decides the outcome. The verdict's soundness rests on the collision-resistance of $H$ and the determinism of the predicate, not on any trusted party.
This paper makes four contributions:
\begin{enumerate}
    \item \textbf{A reframing of the pre-contractual trust deadlock.} We recast confidential bidding not as the problem of verifying a claim before contracting (which the deadlock makes impossible) but as the \emph{optimistic adjudication of committed claims}: a claim is bound at bidding and rendered provably falsifiable under dispute, so that certain exposure under dispute, rather than pre-bid verification or a trusted broker, is what makes the claim usable for pricing.
    \item \textbf{The OVC primitive and a manufacturing pipeline.} We formalize OVC as a commitment, private witness, and deterministic on-chain verification function with precise fault attribution, and compose four instances (Access, Identity, Conformance, and Feature) into a pipeline whose cross-instance consistency is enforced at bind time. All four are implemented in Solidity in both monolithic and segmented execution strategies.
    \item \textbf{The computational boundary of on-chain adjudication.} We empirically characterize the \emph{single-transaction ceiling}: the artifact size beyond which one-shot verification exceeds the per-transaction gas limit of every real chain. Partitioning the work across bounded transactions restores feasibility up to 50~MB artifacts; this segmentation relocates the necessary $O(n)$ work rather than reducing it, and Section~\ref{subsec:scalability} quantifies what it costs.
    \item \textbf{A deployment analysis across the scalability trilemma.} We project the measured gas onto three real chains (Ethereum, Arbitrum, opBNB), quantifying settlement cost and dispute duration at industrial artifact sizes and showing that execution-layer choice, not protocol design, bounds practicality.
\end{enumerate}

The framework's scope is deliberate. Confidentiality is optimistic: it holds on the honest path, and proving a contested claim discloses the witness on-chain, so a counterparty set on forcing disclosure can do so, at which point the Consumer's recourse is to withhold, resolving to \textsc{Withheld} and forgoing the contract to keep the artifact secret. The predicates are intentionally lightweight, each running in $O(n)$ within gas limits to check a specific property (key wrapping, channel integrity, G-code syntax, or a declared physical parameter) rather than general manufacturability. And OVC adjudicates outcomes rather than settling them, leaving the resulting economic and social consequences to the surrounding marketplace. Stronger, disclosure-free confidentiality through zero-knowledge adjudication is a direct extension addressed in future work (Section~\ref{sec:future_work}).

The remainder of this paper is organized as follows. Section~\ref{sec:background} reviews the background of cloud manufacturing and existing blockchain coordination mechanisms. Section~\ref{sec:methods} formalizes the OVC primitive, instantiates four claims across the three pre-contractual phases, and defines the measurement plan. Section~\ref{sec:results} reports that plan in four stages: fixing the operating chunk, then the cost of posting a demand, of binding a Provider, and of adjudicating a dispute. Section~\ref{sec:discussion} discusses the architectural implications, Section~\ref{sec:future_work} outlines future research directions, and Section~\ref{sec:conclusion} concludes.

\section{Background and Problem Positioning}
\label{sec:background}

The protection of intellectual property in decentralized manufacturing lies at the intersection of distributed production architectures, blockchain-based coordination, and cryptographic exchange protocols. Prior work covers each axis independently, but their joint use in open manufacturing markets still leaves a gap between confidentiality and pre-contractual verifiability. This section reviews the current evidence base and positions the unresolved problem addressed by OVC.

\subsection{The Deadlock of Trust in Decentralized Manufacturing}

Decentralized and shared manufacturing paradigms model production as an interoperable service lifecycle \cite{zareiyan_blockchain_2018, yu_blockchain-based_2020-1, lupi_blockchain-based_2023, rozman_scalable_2021}. Table~\ref{tab:lifecycle} formalizes this as seven sequential phases. This work supplies the coordination and verification mechanisms for the first three (Bidding, Binding, and Specification Handover), which together span the entire pre-contractual negotiation, the stage at which the confidentiality--verifiability deadlock arises; the remaining phases place that contribution within the full service lifecycle.

\begin{table}[ht]
\centering
\caption{The decentralized manufacturing service lifecycle.}
\label{tab:lifecycle}
\begin{tabularx}{\textwidth}{cl>{\raggedright\arraybackslash}X}
\toprule
\textbf{\#} & \textbf{Phase} & \textbf{Description} \\ \midrule
1 & Bidding           & The Consumer publishes a manufacturing requirement; interested Providers assess it and quote, and one offer is selected \\
2 & Binding           & Consumer and selected Provider are committed to the agreed terms of the job \\
3 & Specification Handover & The technical specification needed to manufacture the artifact is transferred to the Provider and checked against what was agreed \\
4 & Execution         & Physical manufacturing of the artifact \\
5 & Physical Delivery & Shipment of manufactured artifact to Consumer \\
6 & Validation        & Quality assessment of physical artifact \\
7 & Settlement        & Payment release and contract closure \\ \bottomrule
\end{tabularx}
\end{table}

Blockchain-enhanced cloud manufacturing architectures improve coordination, traceability, and marketplace automation, and a growing body of work has mapped their application across the additive manufacturing domain \cite{grunewald_review_2023}. However, these systems do not eliminate informational asymmetry during bidding, and service pricing in decentralized cloud platforms remains a non-trivial coordination problem \cite{aghamohammadzadeh_novel_2020, yu_blockchain-based_2020, zhu_pricing_2020}.

In additive manufacturing, accurate quotations depend on geometry- and process-derived properties that are tightly coupled to proprietary design assets. Early disclosure of those assets creates direct IP risk, including unauthorized model replication and tampering \cite{holland_intellectual_2018, zhang_using_2022, papakostas_novel_2019}. Industry analyses confirm that IP rights management and data security constitute the primary adoption barriers for blockchain-enabled 3D printing ecosystems \cite{klockner_does_2020, kurpjuweit_blockchain_2021}. This induces the deadlock of trust: Providers require technical evidence to price correctly, while Consumers cannot reveal sensitive files before contractual protection is established, an instance of Arrow's information paradox \cite{arrow_economic_1962}, in which information cannot be valued without being disclosed, nor disclosed without being forfeited.

\subsection{Proactive IP Protection vs. Reactive Quality Assurance}

Two families of countermeasures dominate current practice. The first is proactive access control, including DRM-style streaming, watermark-based authentication, and policy-based encryption. Industry platforms demonstrate encrypted 3D file streaming for print-on-demand networks \cite{secured3d_secured3d:_2022, replique_replique_2023, hubs_how_2024}, while research contributions propose blockchain-enabled authentication frameworks for 3D printing IP \cite{chan_blockchain-enabled_2023}, CP-ABE schemes for fine-grained access control \cite{hu_access_2023, sasikumar_blockchain-assisted_2024, zhang_blockchain-enabled_2024, kryston_martsia:_2025}, and smart-contract-based data confidentiality in P2P industrial communication \cite{stodt_data_2020}. However, these mechanisms do not by themselves prove that the concealed artifact semantically matches the claims used for pricing or capacity reservation.

The second family is reactive process assurance: in-situ monitoring, digital twin synchronization, and sensor-integrated integrity checks. These methods materially improve anomaly detection and post-hoc traceability of print execution \cite{petsiuk_towards_2022, olaoye_-situ_2024, tasneem_hybrid_2024, shi_sensor_2024}. Yet their assurance is predominantly operational-phase assurance; they validate manufacturing behavior during or after production and cannot solve pre-contractual semantic verification.

\subsection{Cryptographic Verification and the Computational Bottleneck}

Fair Exchange and blockchain escrow protocols provide atomic delivery guarantees and stronger dispute semantics for digital assets \cite{asokan_optimistic_2000, tas_atomic_2024, ferrer-gomila_three-step_2026}. Nevertheless, these protocols are fundamentally content-agnostic at verification time: they can confirm key release, payment coupling, or hash consistency, but not whether hidden manufacturing data satisfies functional constraints (e.g., printability or bounded material consumption).

Zero-knowledge systems are the strongest known primitive for proving compliance without revealing raw data, and their application to supply-chain compliance without disclosing proprietary data has been demonstrated \cite{malik_supply_2022}. Recent work reports substantial progress in practical proving pipelines \cite{zhao_zkcnn:_2025, cong_scalable_2025, wang_systematic_2025}. However, current empirical evidence still shows nontrivial proving overhead and implementation complexity that can be difficult to absorb in routine, latency-sensitive industrial bidding workflows \cite{kuznetsov_performance_2025}.

\subsection{Problem Formulation}

Taken together, existing approaches optimize different dimensions of trust but do not jointly provide all three requirements for pre-contractual manufacturing negotiation: (i) confidentiality of proprietary artifacts, (ii) binding exchange guarantees, and (iii) deterministic on-chain adjudication of the concealed artifact's specific claimed properties under dispute. Table \ref{tab:comparison} summarizes this design-space tension.

The open challenge is to make claims over concealed manufacturing artifacts \textbf{binding and provably falsifiable}: participants must be able to commit to, challenge, and adjudicate specific technical and economic claims without requiring immediate disclosure or heavy, monolithic proof circuits for every negotiation. The claim is bound at commitment and rendered exposable under dispute (certain exposure under dispute rather than verification before the contract is formed).

\begin{table}[ht]
\centering
\caption{Comparison of Existing Coordination and Protection Mechanisms in Decentralized Manufacturing}
\label{tab:comparison}
\resizebox{\textwidth}{!}{%
\begin{tabular}{@{}>{\raggedright\arraybackslash}p{3.5cm}>{\raggedright\arraybackslash}p{4cm}cc>{\raggedright\arraybackslash}p{5cm}@{}}
\toprule
\textbf{Approach / Paradigm} & \textbf{Key Mechanism} & \textbf{Pre-Contractual Confidentiality} & \textbf{Semantic Verifiability} & \textbf{Primary Limitation for P2P Bidding} \\ \midrule
\textbf{Policy-Based Confidential Exchange} \cite{hu_access_2023, sasikumar_blockchain-assisted_2024, zhang_blockchain-enabled_2024, kryston_martsia:_2025} & CP-ABE and attribute-governed access over blockchain-assisted workflows. & High & Low & Controls access but does not prove that concealed payload semantics match negotiated claims. \\ \addlinespace
\textbf{DRM / Streaming-Centric IP Control} \cite{secured3d_secured3d:_2022, replique_replique_2023, chan_blockchain-enabled_2023} & Encrypted streaming, license-gated usage, watermarking and blockchain-based authentication. & High & Low & Strong operator control, but central trust dependency and limited trust-minimized dispute semantics. \\ \addlinespace
\textbf{Reactive Process Assurance} \cite{petsiuk_towards_2022, olaoye_-situ_2024, tasneem_hybrid_2024, shi_sensor_2024} & In-situ monitoring, digital twins, sensor-backed anomaly and integrity checks. & Low (Pre-contract) & Medium/High (Post-start) & Evidence arrives during/after execution, too late for pre-contract truthfulness of concealed artifacts. \\ \addlinespace
\textbf{Fair Exchange / Escrow Protocols} \cite{asokan_optimistic_2000, tas_atomic_2024, ferrer-gomila_three-step_2026} & Atomic swap, optimistic arbitration, payment-delivery coupling. & Medium/High & Low & Guarantees delivery fairness, but adjudication remains content-agnostic for manufacturing-specific properties. \\ \addlinespace
\textbf{Zero-Knowledge Verification} \cite{zhao_zkcnn:_2025, cong_scalable_2025, kuznetsov_performance_2025, wang_systematic_2025} & Succinct proof systems for compliance without plaintext disclosure. & High & High & Strong semantics but nontrivial proving cost and integration complexity for routine industrial negotiation. \\ \addlinespace
\textbf{Proposed Framework: OVC} & Segmented claim-aware adjudication with deferred disclosure. & High (Optimistic path) & High (Claim-specific) & Binds claims at bidding and makes them provably falsifiable under dispute, but does not verify them before contracting. \\ \bottomrule
\end{tabular}%
}
\end{table}

\section{Materials and Methods}
\label{sec:methods}

To reconcile the inherent tension between confidentiality and semantic truthfulness in decentralized manufacturing, we introduce the \textbf{Optimistic Verifiable Claim (OVC)} framework. This section formalizes the OVC as a modular primitive that enforces truthful claims without mandating broad intellectual property disclosure.

\subsection{System Model and Adversarial Assumptions}
\label{subsec:model}

The OVC protocol operates within a decentralized marketplace comprising three core entities: the \textit{Consumer} ($\mathcal{C}$), who possesses private design data; the \textit{Provider} ($\mathcal{P}$), who offers manufacturing services; and the \textit{Smart Contract} ($\mathcal{S}$), acting as an automated, code-based arbiter. $\mathcal{S}$ is the \emph{sole executor} of all verification logic: every predicate evaluation, hash check, and feature extraction runs exclusively on-chain, with no reliance on off-chain computation or trusted oracles. Within this environment, an OVC instantiation is defined as a stateful primitive $\Omega = \langle c, w, \mathcal{V} \rangle$, where $c$ is the on-chain commitment, $w$ is the private witness, and $\mathcal{V}$ is the deterministic verification function (formalized in Section~\ref{subsec:atomic_ovc}).

We assume the underlying blockchain guarantees consensus integrity. Participants may behave maliciously to gain informational or economic advantages. The framework addresses the following adversarial behaviors, stated here as the problems to be solved; the protocol's response to each is developed in Section~\ref{subsec:atomic_ovc} and Section~\ref{sec:discussion}:

\begin{itemize}
    \item \textbf{Strategic Misrepresentation}: $\mathcal{C}$ asserts a false claim about the private witness $w$ to obtain favorable contract terms.
    \item \textbf{Witness Tampering}: during a dispute, $\mathcal{C}$ presents a different artifact from the one it originally committed to.
    \item \textbf{IP Probing}: $\mathcal{P}$ initiates a dispute solely to force on-chain disclosure of $w$, independent of the claim's validity.
    \item \textbf{Evidence Withholding}: a party stalls the negotiation by never submitting the verification materials a dispute requires, freezing the contract indefinitely.
\end{itemize}

\subsection{The Atomic OVC Primitive: Formal Definition}
\label{subsec:atomic_ovc}

The Optimistic Verifiable Claim (OVC) primitive resolves the confidentiality–verifiability conflict by enabling \emph{claim-aware} verification. Unlike content-agnostic exchange protocols, an OVC is an atomic unit designed to adjudicate exactly one discrete claim regarding a concealed artifact. A formal OVC instantiation is the stateful primitive governed by the triplet given in Equation~\ref{eq:ovc_triplet}.

\begin{equation}
\Omega = \langle c, w, \mathcal{V} \rangle
\label{eq:ovc_triplet}
\end{equation}

In this formulation, $c$ is the public \textbf{Commitment} published on-chain, $w$ is the \textbf{Private Witness} held by the Consumer ($\mathcal{C}$), and $\mathcal{V}$ is the deterministic \textbf{Verification Function} executed by the Smart Contract ($\mathcal{S}$). To bind the witness to the protocol without revealing its content, $w$ is mapped to $c$ via a collision-resistant cryptographic hash function $H$, as defined in Equation~\ref{eq:commitment_hash}.

\begin{equation}
c = H(w)
\label{eq:commitment_hash}
\end{equation}

\textbf{Commitment properties.} The construction $c = H(w)$ provides \textbf{computational binding}: under the collision-resistance assumption, no efficient adversary can produce $w' \neq w$ such that $H(w') = c$. The construction does \emph{not} provide information-theoretic hiding in the standard commitment-scheme sense; a computationally unbounded party who knows $c$ can in principle brute-force $w$ if the witness space is small. In the OVC protocol, \textbf{pre-dispute confidentiality is a protocol property}, not a cryptographic hiding guarantee: the witness $w$ is never published unless a dispute is explicitly triggered and $\mathcal{C}$ elects to disclose. How far this caveat bites depends on what a given instance commits to, so its per-instance consequences are taken up with the limitations in Section~\ref{subsec:limitations}.

\begin{figure}[H]
    \centering
    \includegraphics[width=0.95\linewidth]{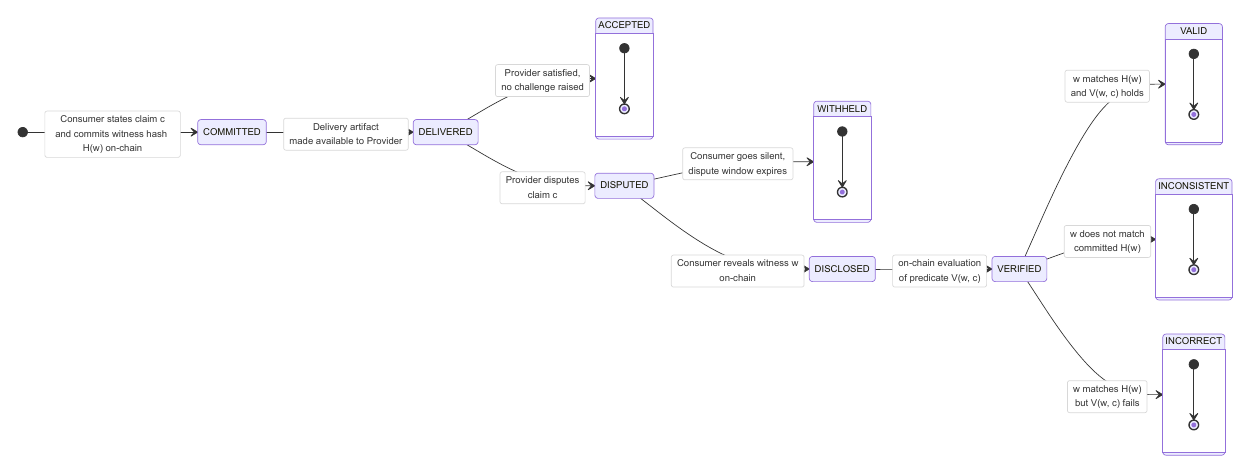}
    \caption{Finite state machine of the atomic OVC primitive.}
    \label{fig:ovc_base}
\end{figure}

The lifecycle of a \emph{single claim} is modeled as a finite-state machine (FSM) in Figure~\ref{fig:ovc_base}. It is distinct from, and nested inside, the seven-phase service lifecycle of Table~\ref{tab:lifecycle}: the state machine below describes one claim from commitment to verdict, and Section~\ref{sec:instantiations} maps its states onto the phases in which they occur. Under optimistic conditions, a claim transitions from \texttt{COMMITTED} to \texttt{ACCEPTED} via a simple acceptance from the Provider ($\mathcal{P}$), requiring no data disclosure. The dispute resolution branch is only triggered if $\mathcal{P}$ issues a challenge, at which point $\mathcal{C}$ must disclose $w$ to the on-chain arbiter $\mathcal{S}$.

The machine also fixes what a claim \emph{is}, which Section~\ref{sec:instantiations} relies on: a claim is not merely a published commitment $c$, but a commitment together with a counterparty empowered to challenge it and a bounded window $W_d$ in which to do so. A commitment with no counterparty is binding on the Consumer yet has no path through this machine, and is therefore not an OVC instance.

The verification function $\mathcal{V}$ evaluates the witness against the specific claim anchored by $c$, as defined in Equation~\ref{eq:v_logic}.

\begin{equation}
\mathcal{V}(w, c) = \mathbb{I}\big(H(w) = c\big) \land \mathcal{P}(w, c)
\label{eq:v_logic}
\end{equation}

Here, $\mathbb{I}(\cdot)$ ensures cryptographic consistency, while $\mathcal{P}(w, c)$ represents the \textbf{Semantic Predicate} for the specific property under test. This modular structure ensures that the terminal outcomes, defined in Equation~\ref{eq:outcomes}, provide precise fault attribution for each individual claim.

\begin{equation}
\text{Outcome} =
\begin{cases}
\text{VALID}             & \text{if } \mathcal{V}(w, c) = 1 \\
\text{INCONSISTENT}      & \text{if } H(w) \neq c \\
\text{INCORRECT}         & \text{if } H(w) = c \land \mathcal{P}(w, c) = 0 \\
\text{WITHHELD}          & \text{if } w \text{ not disclosed within } W_d
\end{cases}
\label{eq:outcomes}
\end{equation}

\textbf{Dispute window $W_d$.} To guarantee unconditional liveness, the protocol includes a configurable dispute window $W_d$, measured in blocks. When a challenge is raised, $W_d$ is set on-chain; if the Consumer has not disclosed the witness by block $b_{challenge} + W_d$, any party may call the contract to resolve the outcome as \textsc{Withheld}. The appropriate value of $W_d$ depends on the target blockchain's block time, the artifact size being adjudicated, and the service-level agreement (SLA) requirements of the service: for fast-settling L2 networks it may be set to a few thousand blocks, while for services requiring extended deliberation on L1 it may span tens of thousands of blocks. The window is a deployment parameter and does not affect the verification logic itself.

By isolating claims into these atomic instances, the protocol enables a "trust-but-verify" mechanism that minimizes intellectual property exposure while ensuring that any semantic misrepresentation results in a provable, on-chain fault.

\subsection{Modular OVC Instantiations for Manufacturing}
\label{sec:instantiations}

The OVC framework is instantiated as a sequential trust chain composed of four claim-specific verifiers. All instantiations share the same atomic OVC state machine (Figure~\ref{fig:ovc_base}) but differ in their \textbf{Private Witness} ($w$), \textbf{Commitment} ($c$), and \textbf{Verification Function} ($\mathcal{V}$). Each instantiation enforces a distinct class of claims relevant to decentralized manufacturing negotiation.

We organize this section by the three pre-contractual phases of the service lifecycle (Table~\ref{tab:lifecycle}), because each phase answers a different question. \textbf{Bidding} asks what is being offered, and is provider-independent: the Consumer publishes a \emph{Demand} so that any Provider can price a job it cannot read. \textbf{Binding} asks who will perform it, and instantiates all four OVC claims against the Provider the Consumer selects. \textbf{Specification Handover} asks whether the claims are true, and is where the four verification functions execute and where any dispute is adjudicated.

A single OVC claim therefore leaves a footprint in more than one phase, and one distinction keeps the phases straight. A \textbf{commitment} is a cryptographic binding to a value; a \textbf{claim} is a commitment together with a named counterparty and a procedure for challenging it. \emph{Bidding produces commitments, binding turns them into claims.} Once the Demand is sealed, $\widetilde{H}(G)$, $f_{claim}$ and $\widetilde{H}(C)$ are public and immutable, and the Consumer is irrevocably bound to them, which is what allows a Provider to price a file it cannot read. Bidding does not supply enforceability: with no Provider selected there is no party holding standing to challenge, and no window in which to do so. Binding supplies both. Specification Handover then resolves the claim, and is where each instance's verification function executes.

The two steps are not always separable in time. The values committed by OVC-Conformance and OVC-Feature describe the artifact alone and are fixed the moment the Demand is posted, whereas those of OVC-Access and OVC-Identity depend on the selected Provider's key material and cannot exist before selection. Binding is therefore the earliest point at which all four can be committed together, which is what permits the single atomic coordinator transaction of Section~\ref{subsubsec:binding_phase}. The four instances are accordingly defined under Specification Handover: what separates OVC-Access from OVC-Feature is not what either commits, since all four commitments are written in that one transaction, but what each one checks when challenged.

\subsubsection{Bidding: The Demand}
\label{subsubsec:demand}

A \textbf{Demand} is the Consumer's public, provider-independent commitment to a specific manufacturing job. It records the artifact commitment $\widetilde{H}(G)$, the publicly asserted feature value $f_{claim}$, and the encrypted artifact $C = G \oplus K$, which is uploaded on-chain so that delivery itself becomes adjudicable. Posting a Demand is what allows competing Providers to price a job they cannot read.

The Demand is a first-class protocol object and its contents are a genuine commitment: once sealed, they cannot be altered, and the Consumer is publicly bound to them. It is not an OVC instance. An OVC claim is defined by its dispute mechanism (Section~\ref{subsec:atomic_ovc}), and a dispute requires an identified counterparty with standing to challenge and a stake in the outcome. At bidding no Provider has yet been selected, so there is no party against whom the assertion could be adjudicated: the Demand carries commitments but no counterparty, no challenge window, and no terminal state. It binds the Consumer to an artifact and to the values asserted about it; it does not adjudicate them.

Adjudication becomes possible only once a counterparty exists. All four OVC instances are therefore instantiated at binding, and each adjudicates values that were fixed earlier, at bidding: OVC-Conformance and OVC-Feature against the Demand's $\widetilde{H}(G)$ and $f_{claim}$, OVC-Identity against the Demand's ciphertext $C$, and OVC-Access against the key material that makes $C$ readable. The separation is what gives the protocol its economic shape: the $O(N)$ artifact upload is paid once at bidding, independently of how many Providers bid, while binding to a chosen Provider is $O(1)$.

\subsubsection{Binding: Instantiating the Four Claims}
\label{subsubsec:binding_phase}

Once the Consumer ($\mathcal{C}$) selects a Provider, the \textbf{Binding phase} executes a single atomic transaction that instantiates all four OVC instances against that Provider and delivers the decryption key. The key $K$ is published on-chain encrypted under the selected Provider's public key, so the Provider can recover $K$, and hence $G = C \oplus K$, entirely from the ledger. Because the ciphertext was uploaded at bidding, binding writes only $O(1)$ state regardless of artifact size. This is the only mandatory on-chain activity of the non-dispute path beyond posting the Demand itself. Table~\ref{tab:binding} summarises what each phase writes to the blockchain.

\begin{table}[H]
\centering
\caption{On-chain state written by the Consumer across the two phases. Bidding posts the provider-independent Demand, including the $O(N)$ ciphertext upload (a single transaction in the monolithic design, chunked in the segmented). Binding instantiates all four OVC instances against the selected Provider and delivers the key; every binding write is $O(1)$.}
\label{tab:binding}
\resizebox{\textwidth}{!}{%
\begin{tabular}{@{}cllll@{}}
\toprule
\textbf{Phase} & \textbf{Object / OVC Instance} & \textbf{Committed on-chain (monolithic)} & \textbf{Committed on-chain (segmented)} & \textbf{Private witness held by Consumer} \\ \midrule
Bidding & Demand (not an OVC instance) & $H(G)$,\; $f_{claim}$,\; full ciphertext $C$ & $\widetilde{H}(G)$,\; $f_{claim}$,\; $C$ (chunked),\; $\widetilde{H}(C)$ & $K$, $G$ \\ \midrule
Binding & OVC-Access      & $H(K)$                                  & $H(K)$                               & $K$ (XOR decryption key) \\
 & OVC-Identity    & $H(K \| G \| C)$                          & $\widetilde{H}(C)$,\; $\widetilde{H}(G)$,\; $H(K)$ & $K$, $G$ \\
 & OVC-Conformance & $H(G)$                                    & $\widetilde{H}(G)$                   & $G$ (plaintext G-code) \\
 & OVC-Feature     & $H(G)$,\; $f_{claim}$                     & $\widetilde{H}(G)$,\; $f_{claim}$    & $G$ (plaintext G-code) \\
 & \multicolumn{4}{l}{\emph{plus} $\mathrm{Enc}_{pk_{\mathcal{P}}}(K)$, the decryption key, encrypted under the selected Provider's public key} \\
\bottomrule
\end{tabular}%
}
\end{table}

The encrypted artifact is uploaded on-chain during the Bidding phase, so the Provider retrieves $C$ directly from the blockchain and recovers $G = C \oplus K$ once $K$ is delivered; no off-chain artifact transfer is required. The monolithic design uploads $C$ in a single transaction, while the segmented design uploads it across bounded transactions; both place $C$ on the common ledger, which is what makes the delivery itself adjudicable. Under both, the rolling hash $\widetilde{H}(C)$ serves the same commitment role as the monolithic single hash, binding the uploaded ciphertext to the committed artifact. OVC-Conformance and OVC-Feature then require no additional delivery: the Provider holds $G$ by the time those instances are evaluated.

The two phases use different cryptography for different reasons. The artifact is protected symmetrically, because $C = G \oplus K$ must be computed once over a multi-megabyte file and re-verified on-chain byte by byte during adjudication. The key is protected asymmetrically, because it must be readable by exactly one Provider, chosen after $C$ was already published. Encrypting a 32-byte key under the Provider's public key costs $O(1)$, which is what keeps binding independent of artifact size.

\paragraph{Cross-instance consistency across the two phases.}
The OVC primitive (Section~\ref{subsec:atomic_ovc}) specifies commitment for a single instance in isolation. When four instances are composed as a pipeline, an additional consistency requirement emerges: the hash $H(K)$ must be identical in OVC-Access (its commitment $c_A$) and in OVC-Identity (its stored providerKeyHash), and the hash $\widetilde{H}(G)$ must be identical across OVC-Identity (auxHash), OVC-Conformance ($c_{conf}$), and OVC-Feature ($c_{feat}$). If each instance were committed in independent transactions, a dishonest Consumer could register a different $\widetilde{H}(G)$ to OVC-Conformance than to OVC-Identity, making a syntactically valid but unrelated G-code file satisfy the conformance check while the identity check references a different artifact. This cross-instance substitution is invisible to per-instance verification.

The binding transaction does not accept the artifact values as arguments. It \emph{reads} $\widetilde{H}(G)$, $\widetilde{H}(C)$ and $f_{claim}$ from the sealed Demand and propagates them identically to all four instances, executing the four \texttt{commitClaimFor} calls within a single atomic \textbf{BindingPhase} coordinator transaction; the shared $H(K)$ passed to OVC-Access and OVC-Identity is likewise written once. A Consumer therefore cannot bind a Provider to an artifact other than the one published for bidding. Cross-instance consistency is a \emph{contract invariant} spanning both phases, rather than an off-chain assumption about Consumer behavior.

\subsubsection{Specification Handover: Verifying the Claims}
\label{subsubsec:handover}

With the artifact published at bidding and the key delivered at binding, the Provider recovers $G = C \oplus K$ and evaluates the four claims against it. This is the phase in which the OVC state machine of Figure~\ref{fig:ovc_base} actually advances: the Provider either accepts, ending the claim optimistically with no adjudication gas, or challenges, whereupon the corresponding verification function is executed on-chain and returns a deterministic verdict with fault attribution.

The four instances are defined below. Each is stated independently (its private witness $w$, its commitment $c$, and its verification function $\mathcal{V}$) so that it can be adopted, replaced, or extended on its own; the pipeline properties that arise from composing them are treated separately in Section~\ref{subsubsec:binding_phase} and Section~\ref{sec:discussion}. They are presented in the order a dispute traverses them, which is also their order of logical dependence: no claim about the artifact's content is meaningful until the delivery that carries it has been shown to be sound.

\paragraph{OVC-Access}
\label{subsubsec:ovc_access}

OVC-Access establishes verifiable delivery readiness and binds it to the authorized counterparty. At binding, the Consumer commits the hash of the XOR decryption key, $c_{A} = H(K)$, together with an auxiliary linkage commitment $c_{aux} = H(K \| H_{prov})$, where $H_{prov}$ is the hash of the authorized Provider's key material. In the same transaction the Consumer publishes $\mathrm{Enc}_{pk_{\mathcal{P}}}(K)$, the key encrypted under the selected Provider's public key, so that key delivery occurs on the ledger rather than over an unobservable side channel: whether the Consumer released the key at all is itself a matter of public record. The verification function checks both the key commitment and the Provider linkage, as shown in Equation~\ref{eq:ovc_access}.

\begin{equation}
\mathcal{V}_{Access}(K,\; c_{A},\; c_{aux}) =
\mathbb{I}\!\big(H(K) = c_{A}\big)
\;\land\;
\big(H(K \| H_{prov}) = c_{aux}\big)
\label{eq:ovc_access}
\end{equation}

\textsc{Inconsistent} is raised when the revealed key does not match $c_A$; \textsc{Incorrect} is raised when the key matches but the auxiliary linkage fails, indicating that the committed key was not bound to the authorized Provider.

\paragraph{OVC-Identity}
\label{subsubsec:ovc_identity}

OVC-Identity enforces the integrity of the delivery channel by binding the decryption key, the plaintext artifact, and the ciphertext into a single cryptographic commitment. The private witness is the triple $w_{ID} = (K, G, C)$, and the commitment is the hash of their concatenation, $c_{ID} = H(K \| G \| C)$. Delivery of $C$ is on-chain in both designs (Table~\ref{tab:binding}): the ciphertext is uploaded during bidding, in a single transaction (monolithic) or across bounded transactions (segmented), with the rolling hash $\widetilde{H}(C)$ serving the same commitment role in the segmented case as the single hash does in the monolithic; in both, the Provider recovers $G = C \oplus K$.

Because $C$ already resides on the ledger as part of the Demand, the instance created at binding does not re-upload it: it references the Demand's ciphertext and adjudicates against it directly, with $\widetilde{H}(C)$ tying the two together. Placing the artifact in the Demand rather than in the instance is what allows one upload to serve every Provider that bids, and it is why the size-proportional cost of the protocol is incurred once, at bidding, rather than once per binding.

The verification function checks two conditions, as shown in Equation~\ref{eq:ovc_identity_abstract}: (i) the revealed triple matches the committed hash; and (ii) the revealed plaintext $G$ is consistent with the on-chain ciphertext $C$ under the XOR key $K$.

\begin{equation}
\mathcal{V}_{Identity}(K, G, C,\; c_{ID}) =
\mathbb{I}\!\big(H(K \| G \| C) = c_{ID}\big)
\;\land\;
\big(G \oplus K = C\big)
\label{eq:ovc_identity_abstract}
\end{equation}

The two-condition structure maps precisely onto the base FSM outcomes: \textsc{Inconsistent} is raised when the revealed triple does not match the committed hash; \textsc{Incorrect} is raised when the hash matches but the XOR consistency fails, indicating that the Consumer committed an internally incoherent triple.

\paragraph{OVC-Conformance}
\label{subsubsec:ovc_conformance}

OVC-Conformance enforces syntactic validity of the committed artifact. The Provider verifies against the delivered plaintext $G$; the primitive is agnostic to how $G$ is delivered. The commitment $c_{conf} = H(G)$ is the artifact hash the Consumer fixed when posting the Demand; the instance that adjudicates it is created at binding, against the selected Provider. The verification function checks that the revealed $G$ matches the commitment and that $G$ is a syntactically valid G-code file, as shown in Equation~\ref{eq:ovc_conformance}.

\begin{equation}
\mathcal{V}_{Conformance}(G,\; c_{conf}) =
\mathbb{I}\!\big(H(G) = c_{conf}\big)
\;\land\;
\text{isGcode}(G)
\label{eq:ovc_conformance}
\end{equation}

In this formulation, $\text{isGcode}(\cdot)$ is a deterministic syntactic validator that checks for compliance with the expected manufacturing file format (e.g., presence of standard toolpath commands G0 and G1). This verifier prevents arbitrary or malformed data from being bound to a bid.

\paragraph{OVC-Feature}
\label{subsubsec:ovc_feature}

OVC-Feature verifies negotiation-relevant features that are deterministically derived from the artifact. As with OVC-Conformance, the Provider verifies against the delivered plaintext $G$, independent of the delivery source. Both the commitment $c_{feat} = H(G)$ and the publicly asserted feature value $f_{claim}$ are taken from the Demand, where they were published so that Providers could price the job; the instance that adjudicates them is created at binding. The verification function checks hash consistency and that the on-chain feature extraction matches the asserted claim, as shown in Equation~\ref{eq:ovc_feature}.

\begin{equation}
\mathcal{V}_{Feature}(G,\; c_{feat},\; f_{claim}) =
\mathbb{I}\!\big(H(G) = c_{feat}\big)
\;\land\;
\big(f(G) = f_{claim}\big)
\label{eq:ovc_feature}
\end{equation}

In this module, $f(\cdot)$ is any feature extraction function meeting exactly three conditions: it must be \emph{deterministic}, so that Consumer and contract derive the same value from the same artifact; \emph{publicly defined}, so that a Provider can price against it before seeing the artifact; and \emph{evaluable on-chain}, so that a contested value can be settled without a trusted party. These conditions are the whole of the instance's requirement on $f$. What the feature happens to measure is immaterial to the construction, which serves a filament-length claim, a machining-time estimate, or a tolerance envelope equally well; any $f$ satisfying the three conditions can be substituted without altering the primitive, its commitment, or its dispute procedure.

In this work $f$ recovers the artifact's net material consumption, the quantity a Provider uses directly in pricing. Its concrete definition for FDM G-code, together with the properties of production slicer output that make it costly to evaluate faithfully, is given with the implementation in Section~\ref{subsubsec:baseline_impl}. Evaluating it costs more than a syntactic scan, and Section~\ref{subsec:scalability} shows this is what makes the semantic predicate, rather than the cryptographic one, the limiting constraint on segmentation. The benefit is that a quantitative claim used during negotiation is derived from unmodified slicer output rather than merely asserted.

\subsection{Operational Scenario: The Conflict of Trust}
\label{subsec:scenario}

\begin{figure}[H]
    \centering
    \includegraphics[width=\textwidth]{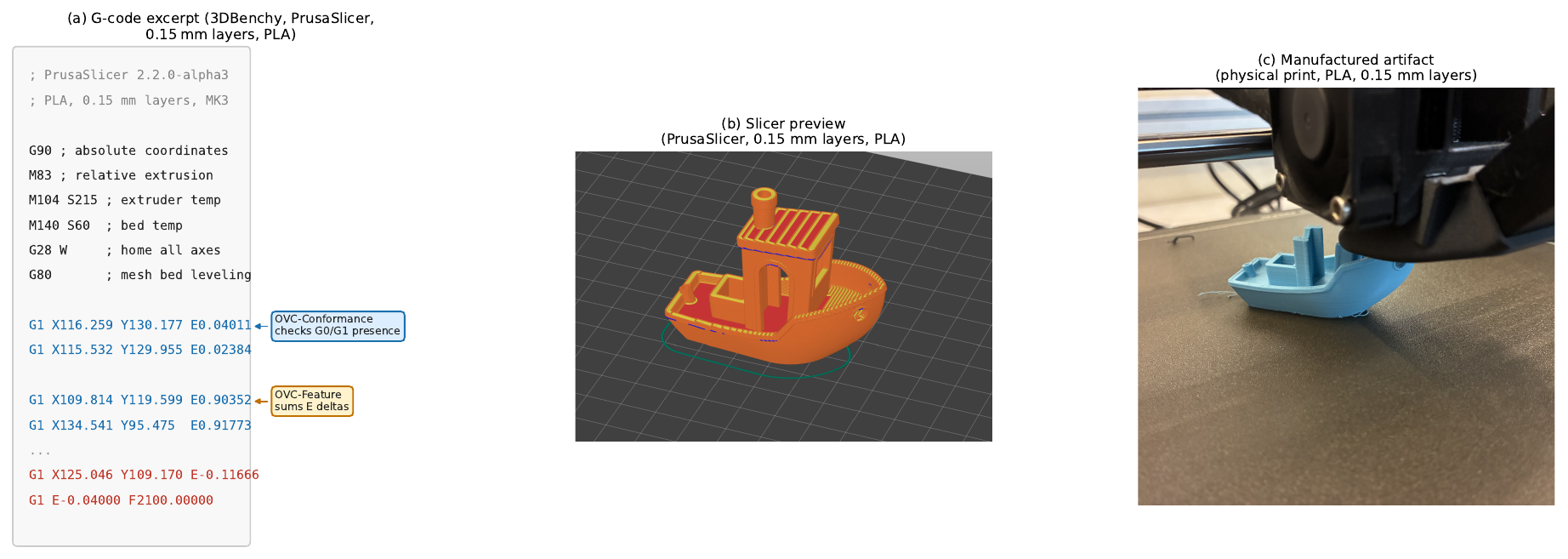}
    \caption{A representative manufacturing artifact (3DBenchy, PLA, 0.15~mm layers, Prusa MK3): \textbf{(a)}~the raw G-code the Consumer holds privately, \textbf{(b)}~its toolpath in top view, layers colored by height, and \textbf{(c)}~the manufactured part.}
    \label{fig:benchy_artifact}
\end{figure}

The utility of the OVC framework is most evident during the \textbf{Bidding phase} of the service lifecycle of Table~\ref{tab:lifecycle}. In a decentralized 3D printing marketplace, a Consumer ($\mathcal{C}$) and a Provider ($\mathcal{P}$) interact to establish the technical parameters of a print job. Figure~\ref{fig:benchy_artifact} illustrates the artifact at the center of this negotiation: a 3DBenchy hull sliced for a Prusa MK3 printer (6.41~MB, 173,351~lines of G-code, PLA at 0.15~mm layer height). Panel~(a) is the raw G-code the Consumer holds privately, over which OVC-Conformance checks for the presence of G0/G1 motion commands and OVC-Feature recovers the declared filament consumption by evaluating the file against a stateful model of the extruder axis (Section~\ref{subsubsec:baseline_impl}); panel~(b) is the toolpath a slicer exposes before printing; panel~(c) is the finished part. On the honest path $\mathcal{C}$ publishes the hash $H(G)$, the declared consumption and the encrypted artifact when posting the Demand at Bidding, releases the decryption key to the selected Provider at Binding, and the job proceeds to Execution without any on-chain dispute. To facilitate the bidding process, $\mathcal{C}$ publishes the \textbf{Demand Tuple} of Equation~\ref{eq:demand_tuple}.

\begin{equation}
\mathcal{D} = \langle c_{conf},\; f_{claim},\; C \rangle
\label{eq:demand_tuple}
\end{equation}

where $c_{conf} = H(G)$ is the hash commitment to the concealed artifact (adjudicated later by OVC-Conformance and OVC-Feature), $f_{claim}$ is the publicly asserted feature value (e.g., total material consumption in mm), and $C = G \oplus K$ is the encrypted artifact, posted on-chain so that delivery is adjudicable rather than assumed. Publishing $C$ discloses nothing without $K$: the geometry, the plaintext G-code, and the decryption key all remain hidden, and $K$ is released only to the Provider selected at binding. Artifact \emph{size} is the one attribute $C$ necessarily reveals, which we return to in Section~\ref{subsec:limitations}. Each interested Provider responds with an \textbf{Offer Tuple}, given in Equation~\ref{eq:offer_tuple}.

\begin{equation}
\mathcal{O} = \langle p_{quote},\; PK_P \rangle
\label{eq:offer_tuple}
\end{equation}

where $p_{quote}$ is the Provider's quoted price derived from $f_{claim}$ and $PK_P$ is the Provider's public key for the encrypted delivery channel. The Provider decides to accept or challenge solely from $\mathcal{D}$: an implausible $f_{claim}$ (for instance, an unusually low material claim for a design of the stated complexity) may rationally trigger a post-delivery dispute. The protocol does not prescribe the challenge rule; it only guarantees that any challenge is deterministically resolved, and the demand can be extended with further OVC instances to cover richer negotiation metadata. In the presented scenario, following the calibration square benchmarked in Section~\ref{subsubsec:baseline_impl}, $\mathcal{C}$ asserts that the concealed design is a functional G-code file requiring exactly $6.0\,\mathrm{mm}$ of filament.

At this juncture, a \textbf{deadlock of trust} arises. Since the marketplace lacks a central authority to verify these claims, $\mathcal{C}$ is unwilling to disclose the raw G-code or the decryption key to protect their intellectual property. Simultaneously, $\mathcal{P}$ is hesitant to conclude the negotiation or dedicate machine capacity without assurance that the technical claims are accurate. Traditional Fair Exchange mechanisms only ensure that \textit{some} data is delivered; they cannot verify if the concealed data satisfies the specific technical requirements claimed during the negotiation.

\begin{figure}[H]
    \centering
    \includegraphics[width=0.72\linewidth,height=0.55\textheight,keepaspectratio]{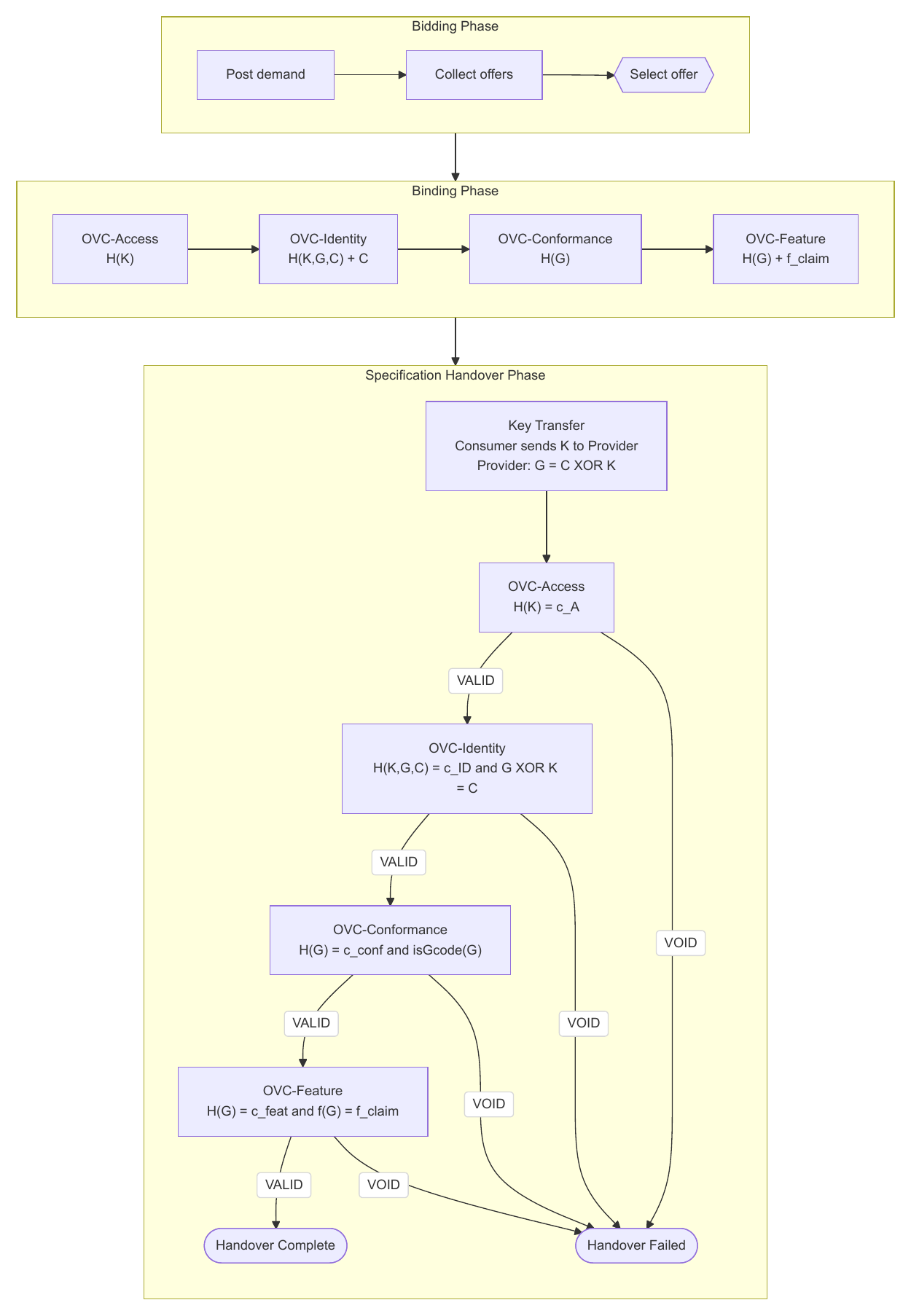}
    \caption{Full OVC protocol experience across the Bidding, Binding, and Specification Handover phases.}
    \label{fig:ovc_pipeline}
\end{figure}

We resolve this by structuring the pre-execution protocol into the three phases of Table~\ref{tab:lifecycle}, whose mechanics are specified in Section~\ref{sec:instantiations} and illustrated end-to-end in Figure~\ref{fig:ovc_pipeline}: $\mathcal{C}$ publishes the Demand at Bidding, binds the selected Provider and releases the encrypted key at Binding, and $\mathcal{P}$ recovers $G = C \oplus K$ and checks it against all four claims at Specification Handover.

That final phase is where the negotiation resolves, in one of two ways. If $\mathcal{P}$ is satisfied it accepts, the phase concludes as \textit{Handover Complete}, and the parties proceed to Execution with no on-chain interaction beyond the Demand and the single binding transaction. If $\mathcal{P}$ disputes, the four instances are adjudicated in dependency order --- OVC-Access $\to$ OVC-Identity $\to$ OVC-Conformance $\to$ OVC-Feature --- each returning a deterministic verdict with precise fault attribution. Should any resolve as other than \textsc{Valid}, the phase concludes as \textit{Handover Failed} and the deal does not proceed.

This staged verification enables the Smart Contract ($\mathcal{S}$) to act as a conditional arbiter, escalating trust only as required. Because the bidding phase anchors the ciphertext before any Provider is selected (the full $C$ on-chain in the monolithic design, its rolling hash $\widetilde{H}(C)$ in the segmented design), and binding propagates that anchor from the sealed Demand rather than accepting it anew, any substitution of the artifact after posting is detectable: a substituted ciphertext fails the on-chain anchor, and a substituted plaintext fails the committed hashes. The artifact a Provider is bound to is therefore provably the artifact it bid on. As a result, the negotiation can conclude with \textbf{computationally sound guarantees} about the validity of all asserted claims, without requiring disclosure of the asset's geometric content unless a dispute is explicitly triggered.

All acceptance and dispute decisions occur during Specification Handover, the third phase, while physical manufacturing starts only in the fourth, Execution. A failed adjudication therefore blocks the job before print-time material or machine costs are incurred; the residual exposure is transaction and opportunity cost, with any monetary compensation delegated to the surrounding marketplace layer.

Worst-case dispute costs for the 3DBenchy artifact, using per-chunk gas measured directly via \texttt{eth\_estimateGas} on the real G-code file, are reported in the bold row of Table~\ref{tab:dispute_all} in Section~\ref{subsec:economics_results}: \$16.86 in 10.8~minutes on opBNB, \$1{,}699 in 15.9~minutes on Arbitrum, and \$42{,}451 in 11.8--47.3~hours on Ethereum (all four instances disputed, beyond the demand publication already paid at bidding). The declared consumption for this artifact, $f(G) = 4{,}020.88$~mm, is recovered on-chain by the OVC-Feature predicate (Section~\ref{subsubsec:baseline_impl}); it exceeds the 3{,}974.5~mm the slicer reports in its own header by 46.4~mm, or 1.2\%. Of that difference, 21.5~mm is the priming extrusion emitted by the start-G-code, which the header excludes; the remainder reflects the slicer's own volumetric accounting, which is not a summation of the $E$ operands it emits. The predicate is defined over the committed artifact, so $f(G)$ is the quantity a Provider can independently recompute and a contract can adjudicate, whereas the header value is advisory metadata.

\subsection{Evaluation Metrics}
\label{subsec:metrics}

Everything reported in this paper reduces to two directly measured quantities: the \textbf{gas} $G$ a transaction consumes, and the \textbf{number of transactions} a job requires. They are not interchangeable. Gas is driven by bytes, because every byte of the witness is hashed, decrypted, or scanned; transaction count is driven by how the witness is divided, because a chunk twice as large costs twice the gas but still occupies a single block. Settlement cost follows from the first, adjudication duration from the second. Both are measured on a high-capacity test network so that raw Ethereum Virtual Machine (EVM) demand is isolated from the block limit of any particular chain; those limits re-enter explicitly, and only, in the projections.

The evaluation proceeds in four stages:

\begin{itemize}
    \item \textbf{R1 --- Chunk selection.} Fixes the operating chunk $c^{*}$ for each deployment chain: the largest chunk whose heaviest transaction still fits within the block gas limit. The monolithic configuration $c = n$ is measured here as the case that fails, which is what motivates segmentation. $c^{*}$ is chosen once and used by every later stage.
    \item \textbf{R2 --- Demand publication.} The cost of publishing a Demand, swept across artifact size at $c^{*}$: gas, settlement cost, and duration. It is $O(n)$ and is incurred \emph{unconditionally} --- before any Provider bids, and whether or not any Provider ever does.
    \item \textbf{R3 --- Binding.} The cost of instantiating the four claims against the selected Provider and having them accepted. $O(1)$ and independent of artifact size, so it is reported as a single row rather than a sweep.
    \item \textbf{R4 --- Dispute adjudication.} The cost of contesting the claims, reported as the \emph{increment} a dispute adds on top of the R2 floor: per instance and for the fully contested case, swept across artifact size at $c^{*}$. It is $O(n)$ and is incurred only if a claim is challenged.
\end{itemize}

Because R1 fixes $c^{*}$, every later stage is a one-dimensional sweep over artifact size; chunk size does not reappear as an axis. The three cost stages differ along two independent dimensions --- how they scale with the artifact ($O(n)$, $O(1)$, $O(n)$) and what triggers them (unconditional, selection, dispute) --- and it is the combination, not the gas figure alone, that determines what each party actually risks.

Per-transaction gas is measured directly at every size on the artifact ladder, including the largest. Where an artifact is split across several chunks, the total is composed arithmetically from the measured per-chunk gas and the chunk count; no multi-thousand-transaction sweep is executed, and every derived figure is identified as such where it is reported.

\subsubsection{Gas and the Cost Model}
The total on-chain gas attributable to one job decomposes into four terms, one per stage, as defined in Equation~\ref{eq:total_gas}.
\begin{equation}
G_{total} = G_{post} + G_{bind} + G_{cha} + G_{adj}
\label{eq:total_gas}
\end{equation}
where $G_{post}$ is the demand-posting cost, dominated by the on-chain artifact upload; $G_{bind}$ the binding commitments and their acceptance; $G_{cha}$ the Provider's challenge; and $G_{adj}$ the on-chain adjudication logic. Two properties of this decomposition carry the economic argument. First, only $G_{post}$ and $G_{adj}$ scale with artifact size; the other two are $O(1)$. Second, the terms are not incurred together: $G_{post}$ is paid whenever a Demand is published, $G_{bind}$ only once a Provider is selected, and $G_{cha} + G_{adj}$ only under dispute. An undisputed job therefore costs $G_{post} + G_{bind}$, of which the first term dominates at industrial scale --- so the cost of an uncontested job is governed not by the protocol's verification machinery but by the size of the artifact it publishes.

\subsubsection{Specific Cost ($\nabla G$)}
To determine how the protocol scales with data volume, we measure the \textbf{specific cost}: the incremental gas demanded per byte of artifact, defined in Equation~\ref{eq:specific_cost}.
\begin{equation}
\nabla G = \frac{\Delta G}{\Delta n}
\label{eq:specific_cost}
\end{equation}
where $n$ is the artifact size in bytes. It applies to both size-proportional stages (the demand upload of R2 and the adjudication of R4) and is the metric that exposes non-linearity: a constant $\nabla G$ indicates cost proportional to artifact size, whereas a rising one signals EVM memory expansion compounding within a single transaction. Comparing $\nabla G$ between the monolithic and segmented configurations is therefore the sharpest test of what segmentation buys.

\subsubsection{Chunk Selection and Feasibility (R1)}
\label{subsubsec:chunk_selection}
A transaction is executable only if its gas fits the block gas limit $G_{limit}$ of the target network. This bounds the artifact sizes a \emph{single} transaction can carry, as defined by the feasibility set in Equation~\ref{eq:feasibility_set}.
\begin{equation}
\mathcal{F}_{real} = \{ n \in \mathbb{N} \mid G(n) \leq G_{limit} \}
\label{eq:feasibility_set}
\end{equation}
Measuring the monolithic configuration, in which the whole artifact is carried in one transaction ($c = n$), locates the boundary of $\mathcal{F}_{real}$ empirically. That boundary --- the \textbf{single-transaction ceiling} --- lies well below industrial artifact sizes, which is the empirical justification for segmentation and the reason the monolithic case appears in this paper only as a comparison.

Segmentation removes the size bound by making the transaction carry a chunk rather than the artifact, so feasibility becomes a constraint on the chunk rather than on the job. The operating chunk is the largest that remains reliably includable, as defined in Equation~\ref{eq:chunk_selection}.
\begin{equation}
c^{*} = \max \left\{ c \;\middle|\; \max_{i} G_{i}(c) \leq \phi \, G_{limit} \right\}
\label{eq:chunk_selection}
\end{equation}
where $G_{i}(c)$ ranges over every per-chunk transaction the protocol issues at chunk size $c$ (the upload and each instance's adjudication), and $\phi \leq 1$ is a fill factor providing headroom below the limit, motivated and quantified in Section~\ref{subsec:scalability}. Taking the maximum over $i$ means one transaction type binds the choice for all of them: deposit and adjudication replay the same chunk sequence under a shared rolling hash, so they cannot be sized independently. Because $c^{*}$ is the largest feasible chunk, it also yields the fewest transactions, and therefore the shortest achievable duration for a given artifact.

\subsubsection{Real-World Settlement Cost ($C_{real}$)}
To evaluate industrial feasibility, the measured gas of any stage is translated into a financial projection. Whereas Equation~\ref{eq:total_gas} divides gas by \emph{stage} --- when it is incurred --- costing requires dividing the same gas by \emph{kind}, since the two components respond differently to artifact size:

\begin{itemize}
    \item \textbf{Intrinsic Gas ($G_{intrinsic}$):} The mandatory overhead for any transaction, consisting of the flat 21,000 gas base fee and the calldata cost of 16 gas per non-zero byte of the witness data. It is what makes a sweep's cost grow with the artifact even before any logic executes.
    \item \textbf{Execution Gas ($G_{exec}$):} The computational resources consumed by the specific OVC verification logic (e.g., parsing, decryption, or feature extraction).
\end{itemize}

The two decompositions are orthogonal views of the same measurement: every stage term of Equation~\ref{eq:total_gas} contains both an intrinsic and an execution component.

The settlement cost of any stage follows from its measured gas, as defined in Equation~\ref{eq:real_cost}.
\begin{equation}
C_{real} = (G_{intrinsic} + G_{exec}) \times P_{gas} \times P_{token}
\label{eq:real_cost}
\end{equation}

where $P_{gas}$ is the effective gas price (base plus priority) and $P_{token}$ the market exchange rate of the chain's native token to USD. Applying it to the terms of Equation~\ref{eq:total_gas} gives the cost of each stage separately, which matters because the stages are triggered under different conditions: the cost a Consumer commits to on publishing a Demand is not the cost of a completed job, and neither is the cost of a contested one.

\subsubsection{Sweep Duration ($T$)}
\label{subsubsec:adjudication_window}

Cost is not the only quantity that scales with artifact size. Both $O(n)$ stages proceed as a \textbf{sweep}: a sequence of chunk transactions, each of which must be confirmed before the next can be issued. A sweep's duration is therefore governed by transaction count rather than by gas, which is why chunk size trades duration against per-transaction weight. It is defined in Equation~\ref{eq:adjudication_window}.
\begin{equation}
T(n) = \left\lceil \frac{n}{c^{*}} \right\rceil \times \beta \cdot \bar{T}_{block}
\label{eq:adjudication_window}
\end{equation}
where $n$ is the artifact size in bytes, $c^{*}$ the operating chunk of Equation~\ref{eq:chunk_selection}, $\bar{T}_{block}$ the mean block confirmation time of the target network, and $\beta \geq 1$ the expected number of blocks between the submission and the inclusion of a transaction. A transaction is not always included in the next produced block even under normal conditions; all projections in this paper use the median inclusion factor $\beta = 1.5$ blocks per transaction. The mean block times used for projection are $\bar{T}_{\mathrm{Ethereum}} = 12$~s, $\bar{T}_{\mathrm{Arbitrum}} = 0.25$~s, and $\bar{T}_{\mathrm{opBNB}} = 1$~s.

The same quantity carries a different operational meaning in each of the two $O(n)$ stages, and the distinction matters for an SLA because only one of them freezes a contract.

\textbf{Time-to-publish (R2).} Applied to the upload sweep, $T$ is the interval between a Consumer beginning to post a Demand and that Demand becoming visible to bidders. It is incurred on the \emph{normal} path --- whether or not any dispute ever occurs, and indeed whether or not any Provider ever bids --- so it delays market entry rather than interrupting an agreed job. It is nonetheless the reason the optimistic path is not instantaneous: a large artifact is not biddable the moment its Consumer decides to publish it.

\textbf{Adjudication window (R4).} Applied to the adjudication sweep, $T$ is the duration during which the service contract is frozen in a contested, unresolvable state, from the first on-chain dispute transaction until the final adjudication transaction is confirmed. During this interval neither party can proceed with contract execution or service delivery, so it maps directly to a \textbf{service availability disruption}. Segmentation also imposes a \textbf{liveness obligation} on the disputing party here: the Consumer ($\mathcal{C}$) must remain online and submit each chunk transaction in sequence within the dispute window $W_d$, and if any chunk is not submitted before $W_d$ expires the contract resolves to \textsc{Withheld}. The adjudication window therefore plays a double role --- it is the SLA-relevant disruption metric, and it is the lower bound for setting $W_d$, which must exceed it by a margin sufficient for network variability and congestion. The remaining lifecycle transactions add at most two further block confirmations, so for industrial-scale artifacts the total contract-freeze duration is $\approx T$.

\subsubsection{Fee-Priority Trade-off}
\label{subsubsec:fee_priority}

Under the EIP-1559 fee mechanism, the effective gas price of Equation~\ref{eq:real_cost} decomposes as shown in Equation~\ref{eq:eip1559}.
\begin{equation}
P_{gas} = B_{fee} + P_{tip}
\label{eq:eip1559}
\end{equation}
where $B_{fee}$ is the protocol-determined base fee (burned upon inclusion) and $P_{tip}$ is the priority tip paid to the block proposer. Validators preferentially order transactions by $P_{tip}$, establishing a direct \textbf{fee-priority trade-off}: a higher tip reduces the expected waiting time to the next block, while a sub-standard tip risks multi-block exclusion under congestion. Under congestion with a delay factor $\delta \geq 1$ scaling the median-inclusion baseline, the effective adjudication window becomes that of Equation~\ref{eq:adjudication_window_congested}.
\begin{equation}
T^{\delta}(n) = \left\lceil \frac{n}{c^{*}} \right\rceil \times \beta \, \delta \cdot \bar{T}_{block}
\label{eq:adjudication_window_congested}
\end{equation}
Because a segmented sweep requires $\lceil n/c^{*} \rceil$ \emph{sequentially dependent} transactions, even a modest delay factor amplifies the window linearly with artifact size, extending the service contract freeze proportionally. This creates a \textbf{cost--availability coupling}: to keep $\delta$ near 1 and minimize the service disruption period, a participant must increase $P_{tip}$, directly raising $C_{real}$. This coupling further motivates L2 deployment, where lower base fees leave greater economic headroom to offer competitive priority tips without inflating the total settlement cost, thereby controlling both the financial burden and the contract freeze duration simultaneously.

\textbf{Durations are reported as ceilings.} Two facts fix the direction of the bound. The chunk $c^{*}$ chosen in R1 is the largest feasible one, so it yields the fewest sequentially dependent transactions and hence the shortest attainable duration: no alternative chunking is faster, which removes the objection that a different configuration would have performed better. What remains free to vary is inclusion delay, and it varies only upward, through $\delta$. Every duration in this paper is therefore reported as a band whose lower end is median inclusion ($\delta = 1$) and whose upper end is the congestion envelope; where a single figure is quoted as a service guarantee, it is the upper end, since for a manufacturing SLA the number that matters is the one the job will not exceed. The same construction applies to both size-proportional stages: R2's upload sweep and R4's adjudication sweep differ in the count $\lceil n/c^{*} \rceil$ and in how many sweeps each performs, not in the form of the bound.

\subsection{Experimental Setup and Implementation}
\label{subsec:setup}

The objective of the experimental phase is to characterize the resource requirements of the OVC framework in an unconstrained environment to inform real-world segmentation strategies.

\subsubsection{Baseline OVC Implementations}
\label{subsubsec:baseline_impl}

We define four primary OVC modules, each targeting a distinct manufacturing or cryptographic property. These modules serve as the technical foundation for the adjudication logic and are implemented as extensions of the \textbf{OVCBase} contract. All verification logic (including \texttt{isGcode}, feature extraction, and XOR-decryption) is executed exclusively by the Smart Contract ($\mathcal{S}$) during dispute adjudication, with no reliance on off-chain computation or trusted oracles.

Each module realizes the verification function defined in Section~\ref{sec:instantiations} (Equations~\eqref{eq:ovc_access}--\eqref{eq:ovc_feature}) as a Solidity extension of \textbf{OVCBase}. We record here only the implementation choices that shape the measured gas:

\begin{itemize}
    \item \textbf{OVC-Access} concatenates the revealed key with the stored Provider key hash via \texttt{abi.encodePacked} before the single Keccak-256 linkage check.
    \item \textbf{OVC-Identity} applies the 32-byte key cyclically, $G_i \oplus K_{(i \bmod 32)} = C_i$. Its demand-posting upload is the framework's only $O(N)$ on-chain write: the monolithic design uploads the full ciphertext in a single transaction ($\approx$711 gas/byte), crossing the single-transaction ceiling beyond $\approx$26~KB, whereas the segmented design uploads it across bounded transactions (feasible to 50~MB); the $O(1)$ binding commitments (including the rolling hashes $\widetilde{H}(C),\widetilde{H}(G)$) add $\approx$167--184\,K gas. Because $C$ resides on-chain from bidding, delivery is guaranteed and adjudicable; only the plaintext $G$ is streamed additionally under dispute, for the OVC-Conformance and OVC-Feature predicates.
    \item \textbf{OVC-Conformance} executes the G0/G1 check as a full linear traversal of the witness byte array, providing the Specific Adjudication Cost ($\nabla G$) baseline for industrial-scale scanning.
    \item \textbf{OVC-Feature} instantiates the feature function $f$ of Section~\ref{subsubsec:ovc_feature} as net material consumption, the sum of extrusion increments over the artifact's motion commands, as shown in Equation~\ref{eq:material_sum}.
    \begin{equation}
    f(G) = \sum_{i \,\in\, \mathcal{M}(G)} \Delta E_i
    \label{eq:material_sum}
    \end{equation}
    where $\mathcal{M}(G)$ is the ordered set of motion commands in $G$ and $\Delta E_i$ is the extrusion increment each contributes. Recovering each $\Delta E_i$ requires interpreting $G$ as a machine program, because four properties of production G-code each break a naive byte scan: (i) extrusion mode is stateful (\texttt{M82} absolute, \texttt{M83} relative, switchable mid-file); (ii) \texttt{G92~E$v$} re-datums the axis without extruding (stock profiles emit thousands --- 2{,}710 for the evaluation model in absolute mode --- so the final $E$ is not the cumulative total, the axis never exceeding 43.84~mm over a 4{,}094~mm print); (iii) increments are signed, so retractions and their re-primes cancel; and (iv) \texttt{E} also parameterises non-motion commands such as \texttt{M907~E538} (motor current), which a context-free scan would misread. The routine tracks mode and datum across the witness with a dedicated helper; values are held as fixed-point integers scaled by $10^5$, matching the five-decimal precision slicers emit. The calibration-square witness used throughout fixes $f_{claim} = 6.0$~mm (integer 600{,}000).
\end{itemize}

\subsubsection{Experimental Environment}
The experiments were conducted using an integrated development stack designed for high-throughput gas tracking and unconstrained EVM execution (Table~\ref{tab:stack}). The Solidity compiler was used with its default settings: the optimizer is \textbf{disabled} (equivalent to \texttt{optimizer.enabled = false}, 0 optimization runs). All reported gas figures therefore represent \textbf{conservative upper bounds}; enabling the IR-based optimizer in a production deployment would reduce gas consumption, widening the already favorable cost margins reported in Section~\ref{subsec:economics_results}.

\begin{table}[ht]
\centering
\caption{Experimental Software Stack}
\label{tab:stack}
\resizebox{\textwidth}{!}{%
\begin{tabular}{lll>{\raggedright\arraybackslash}p{5cm}}
\toprule
\textbf{Component} & \textbf{Software/Tool} & \textbf{Version} & \textbf{Role in Research} \\ \midrule
Local Node & Anvil & v0.2.0+ & Benchmarking unconstrained execution and gas profiling. \\
Framework & Hardhat & v2.x & Compilation and test orchestration for smart contracts. \\
Language & Solidity & v0.8.20 & Implementation of core OVC protocol logic. \\
Optimizer & solc (default) & disabled & Optimizer disabled (0 runs); all reported gas figures are conservative upper bounds. \\
Library & Ethers.js & v6.x & Provider interaction and precise gas measurement. \\
Environment & Node.js & v20.x & Generation of multi-megabyte synthetic data buffers. \\
\bottomrule
\end{tabular}%
}
\end{table}

\subsubsection{Scalability Stress Testing}
To characterize the \textbf{Monolithic Adjudication Profile}, we generated synthetic witnesses spanning the spectrum of industrial 3D printing complexity (Table~\ref{tab:scalability}). Each synthetic witness is constructed as its predicate's worst case: the OVC-Conformance witness places its first motion command at the end of each chunk, so the syntax scan can never terminate early, and the OVC-Feature witness consists entirely of extrusion lines (\texttt{G1 X.. Y.. E..}), so the machine model evaluates state on every line; OVC-Identity's decryption and hashing are content-independent. Costs projected from these witnesses are therefore conservative upper bounds for real slicer output, and the 3DBenchy measurements of Section~\ref{subsec:stage_profiles} quantify the margin. These sizes serve as the basis for calculating the \textbf{Specific Adjudication Cost} ($\nabla G$). For segmented execution, each artifact is additionally processed across a chunk-size spectrum from 1~KB up to the artifact's own size; this sweep locates the optimal operating chunk and supplies the per-chunk gas from which the R1 chunk selection is derived (Table~\ref{tab:chunkselect}).

\begin{table}[ht]
\centering
\caption{Benchmark Artifact Categories with Estimated Complexity}
\label{tab:scalability}
\resizebox{\textwidth}{!}{%
\begin{tabular}{llrr>{\raggedright\arraybackslash}p{3.8cm}}
\toprule
\textbf{Category} & \textbf{Label} & \textbf{Byte Count} & \textbf{Approx. Lines} & \textbf{Context} \\ \midrule
Small Prints & 64 KB & 65,536 & ~2,000 & Calibration cubes. \\
 & 256 KB & 262,144 & ~8,000 & Hobbyist parts. \\
 & 1 MB & 1,048,576 & ~32,000 & Functional prototypes. \\ \midrule
Large Prints & 5 MB & 5,242,880 & ~160,000 & Assembly components. \\
 & \textbf{3DBenchy} & \textbf{6,726,610} & \textbf{173,351} & \textbf{Real benchmark artifact (this paper); PLA, Prusa MK3, 0.15~mm layers.} \\
 & 10 MB & 10,485,760 & ~320,000 & Surgical guides/jigs. \\ \midrule
Industrial & 25 MB & 26,214,400 & ~800,000 & Aerospace parts. \\
 & 50 MB & 52,428,800 & ~1,600,000 & Industrial toolpaths. \\
\bottomrule
\end{tabular}%
}
\end{table}

\subsubsection{Network Economics and Deployment Chains}

To project the \textbf{Real-World Settlement Cost} ($C_{real}$), we apply the measured gas to three real chains that span the scalability trilemma~\cite{reno_navigating_2025}: Ethereum mainnet (L1), Arbitrum One (a standard optimistic rollup, L2), and opBNB (a high-throughput rollup, BNB-denominated). Rather than anchor on a single volatile spot price, we use representative token values: ETH ranged \$1{,}388--\$4{,}955 over the prior twelve months, so we adopt \$3{,}000 (a cycle-representative midpoint), and BNB \$600 on the same representative basis. Absolute USD scales linearly with token price; the chain ranking and all feasibility thresholds are invariant to it. Two structural facts, anticipated here and established in Section~\ref{subsec:scalability}, shape the analysis: Ethereum and Arbitrum admit comparable operating chunks because their per-transaction execution limits are similar, so moving L1$\to$L2 buys cost and speed but \emph{not} materially larger chunks, while only opBNB's larger block admits a substantially larger chunk, and it does so by relaxing decentralization. No production chain admits a block large enough to avoid segmentation for industrial artifacts.

\begin{table}[ht]
\centering
\caption{Deployment-chain parameters for economic and time projections. Token is the representative price of the chain's gas token (ETH for Ethereum and Arbitrum, BNB for opBNB); Cost Rate $=$ Gas Price $\times$ Token.}
\label{tab:economics}
\setlength{\tabcolsep}{3pt}
\footnotesize
\begin{tabular}{lrrrrr}
\toprule
\textbf{Chain} & \textbf{Block Gas Limit} & \textbf{Gas Price} & \textbf{Token} & \textbf{Block Time} & \textbf{Cost Rate} \\ \midrule
Ethereum (L1) & 36~M & 0.5~Gwei & \$3{,}000 & 12~s & \$1{,}500/Ggas \\
Arbitrum (L2) & 32~M (exec) & 0.02~Gwei & \$3{,}000 & 0.25~s & \$60/Ggas \\
opBNB (L2) & 200~M & 0.001~Gwei & \$600 & 1~s & \$0.60/Ggas \\
\bottomrule
\end{tabular}
\end{table}

\section{Results}
\label{sec:results}

This section follows the four-stage measurement plan of Section~\ref{subsec:metrics} in
order. Section~\ref{subsec:scalability} establishes what a single transaction can carry
and fixes the operating chunk $c^{*}$; Section~\ref{subsec:floor_results} measures what
publishing a Demand costs; Section~\ref{subsec:handshake_results} measures binding; and
Section~\ref{subsec:stage_profiles} measures adjudication. Because $c^{*}$ is fixed once,
in the first stage, every subsequent result is a one-dimensional sweep over artifact size.

Reporting is uniform across the three cost stages. Each reports gas, which is
chain-agnostic, and then settlement cost and duration, which are the same gas projected
onto the three deployment chains of Table~\ref{tab:economics} using
Equations~\eqref{eq:real_cost} and~\eqref{eq:adjudication_window_congested}. Durations are
ceilings in the sense of Section~\ref{subsec:metrics}: $c^{*}$ already minimizes
transaction count, so only congestion widens them. Ethereum is reported as a band
($\delta \in [1,4]$); the lower-variance rollups are reported at $\delta \approx 1$.

The dispute outcomes of Equation~\eqref{eq:outcomes} yield five terminal states, numbered
here for reference in the tables that follow: \textsc{Accepted} (Outcome~1), \textsc{Valid}
(2), \textsc{Inconsistent} (3), \textsc{Incorrect} (4), and \textsc{Withheld} (5). Their
definitions are given in Section~\ref{subsec:atomic_ovc}. The measurements below benchmark
the four actively executed outcomes (1--4); \textsc{Withheld} is a timeout-mediated
terminal state rather than a predicate-evaluation case, and is discussed separately. All
gas was measured on a local Ethereum-compatible execution environment with elevated block
limits, so that raw EVM demand is isolated from the capacity of any particular chain.

% ============================================================
\subsection{R1 --- Chunk Selection and the Computational Boundary}
\label{subsec:scalability}
% ============================================================

The first stage answers the question that governs every later one: how much artifact can a
single transaction carry? Its answer fixes the operating chunk $c^{*}$, and with it the
transaction count (and therefore the duration) of every size-proportional stage.

A worst-case dispute performs six size-proportional sweeps over the artifact: the
bidding-phase upload of the ciphertext $C$, two dispute-time re-deposits of the plaintext $G$
(one each for OVC-Conformance and OVC-Feature), and three adjudication sweeps. These reduce
to four distinct cost profiles, because all three deposits are the same operation (writing
bytes to chain) and differ only in which contract receives them. OVC-Identity performs no
deposit at all: it reads $C$ directly from the Demand and takes only a 32-byte key as its
uploaded witness, so the instance that verifies delivery is the one that needs no
re-delivery. OVC-Access is $O(1)$ throughout and never touches the artifact.

\subsubsection{Gas Cost Without Segmentation}
\label{subsubsec:unsegmented}

Each size-proportional process was measured as a single transaction spanning the whole
artifact, on a node with raised gas limits so that the measurement is not truncated by block
capacity. Table~\ref{tab:mono} reports gas consumption across the artifact ladder.

\begin{table}[H]
\centering
\caption{Gas consumed by each process executed as a single transaction over the whole
artifact. The deposit column applies three times in a worst-case dispute: the bidding-phase
upload of $C$ and the two dispute-time re-deposits of $G$ made by OVC-Conformance and
OVC-Feature. OVC-Identity reads $C$ from the Demand and deposits nothing. OVC-Access is
$O(1)$ and does not scale with the artifact.}
\label{tab:mono}
\resizebox{\textwidth}{!}{%
\begin{tabular}{lrrrrr}
\toprule
\textbf{Artifact} & \textbf{Deposit / upload} & \textbf{OVC-Identity} & \textbf{OVC-Conformance} & \textbf{OVC-Feature} & \textbf{OVC-Access} \\
\midrule
1~KB    & 788{,}716            & 1{,}428{,}918        & 518{,}576            & 1{,}251{,}978        & 63{,}971 \\
4~KB    & 2{,}969{,}677        & 5{,}595{,}443        & 1{,}963{,}600        & 4{,}896{,}450        & 63{,}971 \\
16~KB   & 11{,}693{,}880       & 22{,}262{,}981       & 7{,}744{,}056        & 19{,}461{,}446       & 63{,}971 \\
64~KB   & 46{,}596{,}450       & 88{,}956{,}173       & 30{,}871{,}638       & 77{,}740{,}440       & 63{,}971 \\
256~KB  & 186{,}298{,}890      & 356{,}097{,}581      & 123{,}474{,}126      & 310{,}955{,}202      & 63{,}971 \\
512~KB  & 372{,}798{,}186      & 713{,}203{,}629      & 247{,}173{,}486      & 622{,}137{,}594      & 63{,}971 \\
1~MB    & 746{,}583{,}210      & 1{,}430{,}561{,}453  & 495{,}358{,}638      & 1{,}245{,}282{,}184  & 63{,}971 \\
5~MB    & 3{,}774{,}612{,}138  & 7{,}320{,}418{,}989  & 2{,}518{,}588{,}590  & 6{,}268{,}200{,}892  & 63{,}971 \\
10~MB   & 7{,}654{,}020{,}138  & 15{,}060{,}228{,}281 & 5{,}141{,}997{,}870  & 12{,}641{,}224{,}430 & 63{,}971 \\
25~MB   & 19{,}921{,}389{,}750 & 40{,}796{,}238{,}521 & 13{,}641{,}371{,}322 & 32{,}389{,}434{,}030 & 63{,}971 \\
50~MB   & 42{,}464{,}157{,}750 & 92{,}078{,}196{,}921 & 29{,}904{,}145{,}722 & 67{,}400{,}273{,}094 & 63{,}971 \\
\bottomrule
\end{tabular}%
}
\end{table}

Cost grows linearly with the artifact in every column, spanning four orders of magnitude
between the 1~KB and 50~MB cases. OVC-Access is the exception: its witness is a 32-byte key,
so its cost is constant at 63{,}971 gas and independent of what is being manufactured.

\begin{figure}[H]
    \centering
    \includegraphics[width=0.92\textwidth]{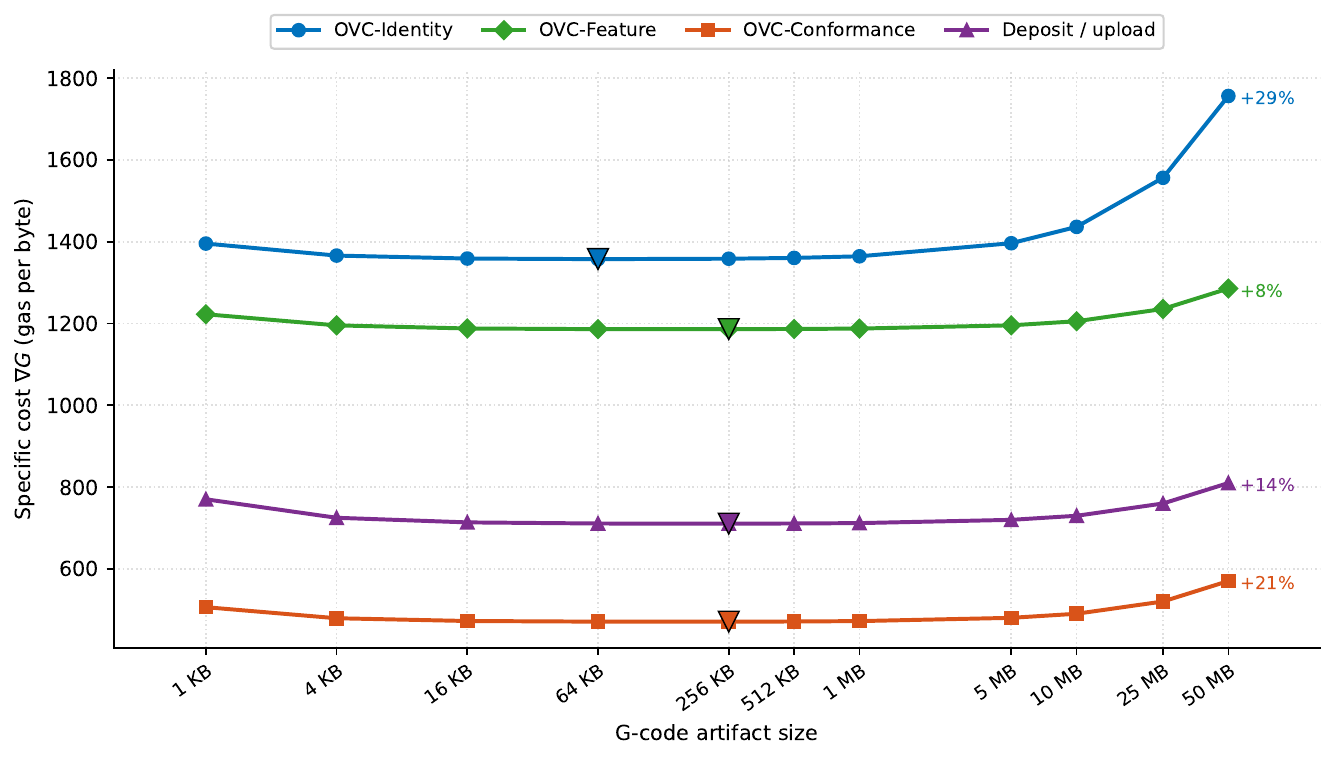}
    \caption{Specific cost $\nabla G$ against artifact size, without segmentation. Markers
    locate each process's minimum; the annotation gives the rise from that minimum to 50~MB.
    The falling arm is fixed per-transaction overhead being amortised; the rising arm is EVM
    memory expansion within a single transaction.}
    \label{fig:specific_cost}
\end{figure}

Specific cost is not constant with artifact size. Figure~\ref{fig:specific_cost} plots
$\nabla G$ (Equation~\ref{eq:specific_cost}) for each process across the ladder, and every
curve is U-shaped. The two arms have different causes. Below roughly 64~KB the fixed
per-transaction overhead (the 21{,}000 gas base charge and the calldata cost of the
witness) is amortised over too few bytes, so specific cost falls as the artifact grows.
Above roughly 256~KB the quadratic EVM memory-expansion term takes over and it rises again.

The rising arm is the one that matters for deployment. Measured from each process's own
minimum to 50~MB, specific cost increases by 29\% for OVC-Identity, 21\% for
OVC-Conformance, 14\% for the deposit and 8\% for OVC-Feature. The instances that hold the
largest working set in memory degrade fastest: OVC-Identity materializes the full triple
$K \| G \| C$, whereas OVC-Feature carries only its accumulator and datum state. Executing a
process in one transaction therefore becomes progressively less efficient as the artifact
grows.

\subsubsection{Segmented Execution and Per-Chunk Cost}

Under segmentation each transaction carries one chunk of the artifact rather than the whole
of it, and EVM memory is released between transactions. The total quantity of work is
unchanged; what changes is how it is divided, and therefore what a single transaction must
hold. Chunk size becomes the free parameter, and the measurements below characterize its
effect on each process independently.

\begin{figure}[H]
    \centering
    \includegraphics[width=0.92\textwidth]{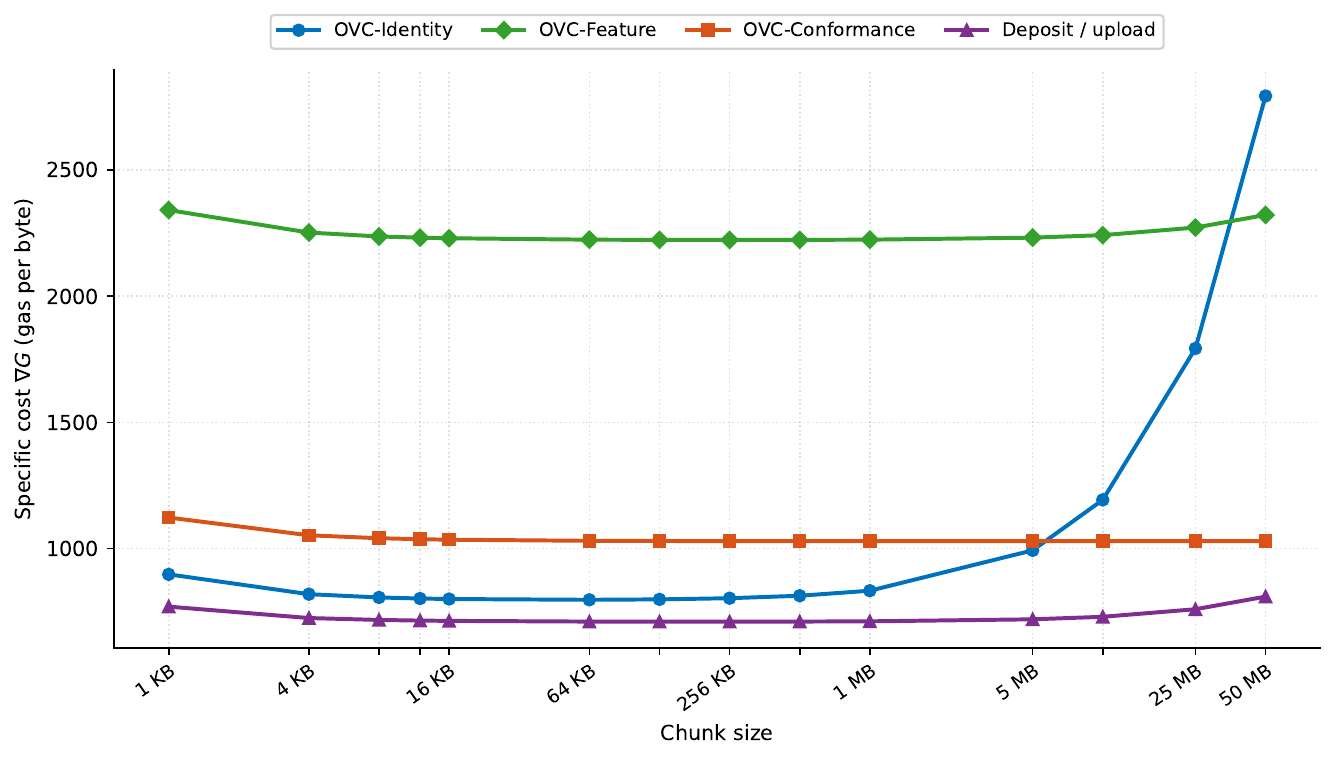}
    \caption{Specific cost $\nabla G$ against chunk size under segmentation. All points are
    measured; the larger chunk sizes require a node with raised gas limits.}
    \label{fig:chunk_cost}
\end{figure}

Figure~\ref{fig:chunk_cost} plots specific cost against chunk size for the four
size-proportional processes, the same quantity
Figure~\ref{fig:specific_cost} plots against \emph{artifact} size
without segmentation; read together the pair states what segmentation accomplishes, in that
cost ceases to depend on the size of the artifact and comes instead to depend on the size of
the chunk.

Table~\ref{tab:chunkgas} gives the corresponding per-chunk transaction gas at each measured
chunk size.

\begin{table}[H]
\centering
\caption{Per-chunk transaction gas by process and chunk size. Each figure is the gas
consumed by one transaction carrying one chunk. The deposit column applies to the bidding-phase upload of $C$ and to the two
dispute-time re-deposits of $G$.}
\label{tab:chunkgas}
\resizebox{\textwidth}{!}{%
\begin{tabular}{lrrrr}
\toprule
\textbf{Chunk} & \textbf{Deposit / upload} & \textbf{OVC-Identity} & \textbf{OVC-Conformance} & \textbf{OVC-Feature} \\
\midrule
1~KB    & 0.79~M   & 0.92~M   & 1.15~M   & 2.40~M \\
4~KB    & 2.97~M   & 3.36~M   & 4.31~M   & 9.23~M \\
8~KB    & 5.88~M   & 6.61~M   & 8.53~M   & 18.32~M \\
12~KB   & 8.79~M   & 9.86~M   & 12.75~M  & 27.43~M \\
16~KB   & 11.69~M  & 13.11~M  & 16.96~M  & 36.53~M \\
64~KB   & 46.60~M  & 52.25~M  & 67.57~M  & 145.75~M \\
128~KB  & 93.15~M  & 104.72~M & 135.03~M & 291.39~M \\
256~KB  & 186.30~M & 210.65~M & 269.97~M & 582.71~M \\
\bottomrule
\end{tabular}%
}
\end{table}

Two features of these measurements are worth noting.

First, specific cost is close to constant over most of the measured range. Between 1~KB and
128~KB every process remains within 14\% of its own minimum, and between 4~KB and 128~KB
within 3\%. Chunk size is therefore not an instrument for reducing gas.

Second, the processes diverge sharply at large chunk sizes, and the divergence separates
them by how much of the chunk each holds in memory. Beyond approximately 1~MB OVC-Identity
rises steeply, reaching 2{,}793 gas/byte at 50~MB against 797 at its minimum. Its
\texttt{processChunk} materializes three buffers per chunk (the ciphertext returned by the
Demand, the decrypted copy, and the input to the rolling hash), and the EVM charges memory
quadratically in the highest offset touched: on the caller side alone that accounts for some
900 gas/byte at a 50~MB chunk against 0.5 gas/byte at 14~KB, with the external fetch
expanding memory in the Demand as well. OVC-Conformance reads its chunk from storage and
retains nothing between bytes, so it stays flat at 1{,}030 gas/byte across the same
interval; OVC-Feature and the deposit rise by 4\% and 14\% respectively. The divergence
identifies memory residency, rather than the arithmetic each predicate performs, as the
property that governs behavior at large chunk sizes.

\subsubsection{Chunk Bounds Under Deployment Constraints}
\label{subsubsec:choosing}

The measurements so far are chain-agnostic. A deployment chain imposes the constraint that
gives chunk size an upper bound: no transaction may consume more gas than the block that
carries it. Ethereum admits 36~M gas per block, Arbitrum 32~M of execution gas, and opBNB
200~M (Table~\ref{tab:economics}).

We size a chunk to at most a fill factor $\phi = 0.95$ of that limit rather than to the
limit itself. Two properties of the fee market motivate the margin. A transaction consuming
the entire limit is includable only in an otherwise empty block, so any competing demand
postpones it. And EIP-1559 sets the per-block gas \emph{target} to half the limit via an
elasticity multiplier of two, raising the base fee by up to $1/8$ per block whenever a block
exceeds that target~\cite{buterin_eip1559_2019}; sustained operation at the ceiling drives
the base fee up at precisely the moment a party is obliged to keep
submitting~\cite{roughgarden_transaction_2020}. Empirically, waiting time is governed by this
congestion dynamic rather than by nominal capacity~\cite{liu_empirical_2022}. The value of
$\phi$ is an engineering choice rather than a protocol constant; it is reported explicitly so
that any other choice can be applied to the same measurements.

Applying the resulting budget $\phi \, G_{limit}$ to the per-chunk gas of
Table~\ref{tab:chunkgas} bounds the chunk each process can carry. Table~\ref{tab:chunkbound}
gives those bounds.

\begin{table}[H]
\centering
\caption{Largest chunk each process can carry on each deployment chain, taken independently.
Each entry is the chunk at which that process's per-chunk transaction reaches
$\phi \, G_{limit}$, with $\phi = 0.95$.}
\label{tab:chunkbound}
\begin{tabular}{lrrr}
\toprule
\textbf{Process} & \textbf{Ethereum} & \textbf{Arbitrum} & \textbf{opBNB} \\
 & \small 36~M & \small 32~M & \small 200~M \\
\midrule
Deposit / upload & 46.8~KB & 41.6~KB & 259.8~KB \\
OVC-Identity     & 41.7~KB & 37.1~KB & 231.6~KB \\
OVC-Conformance  & 32.2~KB & 28.7~KB & 179.1~KB \\
OVC-Feature      & 15.0~KB & 13.3~KB & 83.2~KB \\
\bottomrule
\end{tabular}
\end{table}

Taken on its own, each process admits a different chunk: the deposit carries substantially
more than OVC-Feature on every chain. The bounds scale with the block limit but their
relation to one another does not, being set by the processes' relative cost per byte, which
is a property of the predicates rather than of the chain.

\subsubsection{The Operating Chunk}
\label{subsubsec:operating}

The bounds of Table~\ref{tab:chunkbound} cannot be realized independently. A single
rolling-hash commitment fixes the chunk sequence: $\widetilde{H}$ is an ordered fold over the
deposited chunks, so its value depends on where the boundaries fall, and any party that must
reproduce it has to replay those same boundaries. OVC-Conformance and OVC-Feature verify their
re-deposited $G$ against $\widetilde{H}(G)$ before adjudication may begin, and OVC-Identity
accumulates the decrypted ciphertext into that same value as its verdict. All three therefore
inherit the boundaries fixed by the bidding upload, and the operating chunk $c^{*}$ of
Equation~\ref{eq:chunk_selection} is the \emph{minimum} of the four bounds: on every chain
considered here, OVC-Feature's. Table~\ref{tab:chunkselect} gives it.

\begin{table}[H]
\centering
\caption{R1 operating chunk by deployment chain: the largest measured chunk whose OVC-Feature
transaction, the limiting one of Table~\ref{tab:chunkbound}, fits within
$\phi \, G_{limit}$. Fill is that transaction as a fraction of the block gas limit.}
\label{tab:chunkselect}
\begin{tabular}{lrrr}
\toprule
\textbf{Chain} & \textbf{Operating chunk $c^{*}$} & \textbf{OVC-Feature gas} & \textbf{Fill} \\
\midrule
Ethereum (36~M)  & 14~KB & 31.98~M  & 88.8\% \\
Arbitrum (32~M)  & 13~KB & 29.70~M  & 92.8\% \\
opBNB (200~M)    & 80~KB & 182.15~M & 91.1\% \\
\bottomrule
\end{tabular}
\end{table}

Holding every process to OVC-Feature's bound rather than its own carries little gas penalty:
across the feasible 4--80~KB range total gas varies by about 2\%, while the transaction count
over the same range varies twentyfold. The coupling costs latency, not gas. Every cost
reported in the remainder of Section~\ref{sec:results} is measured at these chunks.

\subsection{R2 --- Cost and Duration of Demand Publication}
\label{subsec:floor_results}
% ============================================================

Publishing a Demand is the protocol's only unconditional cost and its only size-proportional
one on the undisputed path: the cost \emph{floor} beneath every job, contested or not. The
ciphertext $C$ is deposited on-chain in chunks at $c^{*}$, at
the per-chunk rate of Table~\ref{tab:chunkgas}, preceded by \texttt{postDemand} and followed
by \texttt{seal}. Table~\ref{tab:floor} projects gas, settlement cost and time to publish
across the artifact ladder.

\begin{table}[H]
\centering
\caption{R2 demand publication across the artifact ladder, at the operating chunks of
Table~\ref{tab:chunkselect}. The 3DBenchy row is the real 6.41~MB artifact; the remainder are synthetic. Gas is quoted at Ethereum's $c^{*}$; the other chains differ by
less than 0.5\%, chunk size having almost no effect on the total. Durations are ceilings in
the sense of Section~\ref{subsec:metrics}: $c^{*}$ already minimizes transaction count, so
only congestion widens them. Ethereum is reported as a band ($\delta \in [1,4]$), the
lower-variance rollups at $\delta \approx 1$.}
\label{tab:floor}
\setlength{\tabcolsep}{3.5pt}
\footnotesize
\begin{tabular}{lrrrrrrr}
\toprule
 & & \multicolumn{2}{c}{\textbf{Ethereum}} & \multicolumn{2}{c}{\textbf{Arbitrum}} & \multicolumn{2}{c}{\textbf{opBNB}} \\
\cmidrule(lr){3-4} \cmidrule(lr){5-6} \cmidrule(lr){7-8}
\textbf{Artifact} & \textbf{Gas (M)} & \textbf{Cost} & \textbf{Time} & \textbf{Cost} & \textbf{Time} & \textbf{Cost} & \textbf{Time} \\
\midrule
64~KB & 47.0 & \$70.45 & 2.1--8.4~min & \$2.82 & 3~s & \$0.0280 & 4~s \\
256~KB & 187.4 & \$281.09 & 6.3--25.2~min & \$11.25 & 8~s & \$0.1119 & 9~s \\
1~MB & 749.1 & \$1{,}124 & 0.4--1.5~h & \$44.97 & 30~s & \$0.4473 & 22~s \\
5~MB & 3{,}745.0 & \$5{,}617 & 1.8--7.4~h & \$224.80 & 2.5~min & \$2.24 & 1.6~min \\
\textbf{3DBenchy} & \textbf{4{,}804.8} & \textbf{\$7{,}207} & \textbf{2.4--9.4~h} & \textbf{\$288.42} & \textbf{3.2~min} & \textbf{\$2.87} & \textbf{2.1~min} \\
10~MB & 7{,}489.8 & \$11{,}235 & 3.7--14.7~h & \$449.60 & 4.9~min & \$4.47 & 3.2~min \\
25~MB & 18{,}724.4 & \$28{,}087 & 9.2--36.6~h & \$1{,}124 & 12.3~min & \$11.18 & 8.1~min \\
50~MB & 37{,}448.6 & \$56{,}173 & 18--73~h & \$2{,}248 & 24.6~min & \$22.36 & 16.1~min \\
\bottomrule
\end{tabular}
\end{table}

Gas is linear in artifact size to within the fixed overhead of the two $O(1)$ transactions:
across the ladder an 800-fold increase in artifact size raises gas 797-fold. Segmentation
does not change what publication costs, only how many transactions it is spread over. Two
further features of the projection bear on deployment.

First, the cost is unconditional. It falls due when the Demand is published, before any
Provider has bid, and whether or not any Provider ever does. A Consumer publishing a 50~MB
demand on Ethereum commits \$56{,}173 to the act of asking; for the 6.41~MB 3DBenchy the
figure is \$7{,}207. The same two acts cost \$2{,}248 and \$288.42 on Arbitrum, and \$22.36
and \$2.87 on opBNB. This is the only stage whose cost a participant cannot avoid by behaving
well, because it follows from publication itself rather than from dispute or from counterparty
conduct.

Second, the duration is of comparable consequence. Posting a 50~MB demand at $c^{*} = 14$~KB
requires 3{,}658 sequentially dependent upload transactions, so the artifact is not visible to
bidders for between 18 and 73~hours depending on congestion; the 3DBenchy takes 2.4--9.4~hours
on the same chain. This is time-to-publish in the sense of
Section~\ref{subsubsec:adjudication_window}, a delay on the \emph{normal} path, not the
exceptional one. It does not freeze an existing contract, because no contract exists yet; it
delays market entry. An SLA that budgets only for dispute latency will therefore mis-model the
protocol, and on Layer~1 the publication delay alone exceeds what many manufacturing
negotiations would tolerate.

The floor is the price of making delivery adjudicable. Because the ciphertext resides on the
common ledger rather than on a channel between two parties, no Consumer can later claim to
have delivered an artifact it did not publish, and no Provider can deny receiving one that it
did. That guarantee is unavailable to any design that moves the artifact off-chain, and it is
what the upload buys. Two properties soften the cost without removing it: one upload serves
every Provider that bids, so the expense amortises across a competitive field rather than
being repeated per offer; and because the Demand is provider-independent, the same upload also
serves a re-binding if a first Provider's claims fail or the deal lapses. The floor is paid
per \emph{artifact}, not per negotiation.

% ============================================================
\subsection{R3 --- Cost of Binding and Claim Instantiation}
\label{subsec:handshake_results}
% ============================================================

Once a Provider is selected, binding instantiates all four OVC claims against it and delivers
the encrypted key, in two transactions: the Consumer's \texttt{bindProvider}, which commits
all four instances atomically and carries the key, and the Provider's \texttt{accept}, which
co-signs them. Table~\ref{tab:handshake} reports the measured cost and duration. The stage
was measured at artifact sizes from 64~KB to 5~MB on a freshly deployed contract set per
size, and its gas is identical at every size: binding writes a fixed number of
fixed-width slots (commitments, hashes, one encrypted key), whatever artifact stands
behind them, so one row covers the whole ladder.

\begin{table}[H]
\centering
\caption{R3 binding cost and duration by deployment chain. The stage is two transactions,
one from each party, and is independent of artifact size: its gas is measured identical from
64~KB to 5~MB. Durations are ceilings in the sense of Section~\ref{subsec:metrics}; Ethereum
is reported as a band ($\delta \in [1,4]$), the lower-variance rollups at
$\delta \approx 1$.}
\label{tab:handshake}
\setlength{\tabcolsep}{3.5pt}
\footnotesize
\begin{tabular}{lrrrrrrr}
\toprule
 & & \multicolumn{2}{c}{\textbf{Ethereum}} & \multicolumn{2}{c}{\textbf{Arbitrum}} & \multicolumn{2}{c}{\textbf{opBNB}} \\
\cmidrule(lr){3-4} \cmidrule(lr){5-6} \cmidrule(lr){7-8}
\textbf{Stage} & \textbf{Gas (M)} & \textbf{Cost} & \textbf{Time} & \textbf{Cost} & \textbf{Time} & \textbf{Cost} & \textbf{Time} \\
\midrule
Binding & 1.0 & \$1.51 & 0.6--2.4~min & \$0.0605 & 1~s & \$0.0006 & 3~s \\
\bottomrule
\end{tabular}
\end{table}

The handshake costs 1{,}007{,}554 gas in total, whatever the artifact. Set against its
neighbors, the stage is negligible: about 2\% of publishing even the 64~KB artifact of
Table~\ref{tab:floor} on the same chain, and 0.02\% of publishing the 3DBenchy; the ratio
is the same on every chain, being a ratio of gas alone. Binding is nevertheless where all
four claims are created, where every cross-instance invariant is enforced
(Section~\ref{subsubsec:binding_phase}), and where the key is delivered. What the protocol
charges for is not committing to the artifact but publishing it
(Section~\ref{subsec:floor_results}) and, if challenged, re-executing it on-chain
(Section~\ref{subsec:stage_profiles}). The commitment apparatus that makes both adjudicable
is close to free; the data movement around it is not.

% ============================================================
\subsection{R4 --- Cost and Duration of Dispute Adjudication}
\label{subsec:stage_profiles}
% ============================================================

The final stage measures what a dispute costs. It is the only stage incurred conditionally:
nothing here is paid unless a Provider actually challenges a claim, and every figure in this
section is the \emph{increment} a dispute adds on top of the floor of
Table~\ref{tab:floor}, which is paid at bidding whether or not a dispute ever occurs.
Disputes are raised per instance, so the four instances are reported separately, in
ascending order of cost, followed by the worst case in which all four are contested at
once. Each size-proportional instance is reported across the artifact ladder in the format
of Table~\ref{tab:floor}; in every table the 3DBenchy row substitutes per-chunk gas
measured on the real artifact for the synthetic rate, durations count the sweeps plus the
$O(1)$ dispute transactions, gas is quoted at Ethereum's $c^{*}$ (the other chains differ
by less than 0.7\%), and Ethereum is reported as a band ($\delta \in [1,4]$), the
lower-variance rollups at $\delta \approx 1$.

\subsubsection{OVC-Access}

A dispute of the key-escrow claim is the degenerate case: its witness is the 32-byte key,
so adjudication is $O(1)$. The measured increment is 63{,}971 gas (\$0.10 on Ethereum,
\$0.004 on Arbitrum, \$0.00004 on opBNB), independent of artifact size, and settles in a
handful of transactions. It is cheaper than the binding handshake itself and does not
appear again below; the three remaining instances are the size-proportional ones.

\subsubsection{OVC-Identity}

Contesting delivery-channel integrity is the cheapest size-proportional dispute:
adjudication is a single sweep of the ciphertext already resident in the Demand, with no
deposit of $G$, proceeding at roughly 0.80~k gas/byte at the operating chunk
(Table~\ref{tab:dispute_identity}). The 3DBenchy row lands within 0.2\% of the synthetic
rate, because XOR decryption and rolling-hash recomputation are content-independent: the
predicate costs the same whatever the bytes are.

\begin{table}[H]
\centering
\caption{R4 dispute increment for OVC-Identity: one adjudication sweep of the ciphertext in
the Demand, no deposit of $G$.}
\label{tab:dispute_identity}
\setlength{\tabcolsep}{3pt}
\footnotesize
\begin{tabular}{lrrrrrrr}
\toprule
 & & \multicolumn{2}{c}{\textbf{Ethereum}} & \multicolumn{2}{c}{\textbf{Arbitrum}} & \multicolumn{2}{c}{\textbf{opBNB}} \\
\cmidrule(lr){3-4} \cmidrule(lr){5-6} \cmidrule(lr){7-8}
\textbf{Artifact} & \textbf{Gas (G)} & \textbf{Cost} & \textbf{Time} & \textbf{Cost} & \textbf{Time} & \textbf{Cost} & \textbf{Time} \\
\midrule
64~KB & 0.05 & \$78.76 & 2.7--10.8~min & \$3.15 & 3~s & \$0.0314 & 8~s \\
256~KB & 0.21 & \$315.05 & 6.9--27.6~min & \$12.61 & 9~s & \$0.1255 & 12~s \\
1~MB & 0.84 & \$1{,}260 & 0.4--1.6~h & \$50.44 & 31~s & \$0.5018 & 26~s \\
5~MB & 4.20 & \$6{,}301 & 1.9--7.4~h & \$252.21 & 2.5~min & \$2.51 & 1.7~min \\
\textbf{3DBenchy} & \textbf{5.40} & \textbf{\$8{,}098} & \textbf{2.4--9.5~h} & \textbf{\$324.02} & \textbf{3.2~min} & \textbf{\$3.25} & \textbf{2.2~min} \\
10~MB & 8.40 & \$12{,}602 & 3.7--14.7~h & \$504.41 & 5.0~min & \$5.02 & 3.3~min \\
25~MB & 21.00 & \$31{,}505 & 9.2--36.7~h & \$1{,}261 & 12.3~min & \$12.55 & 8.1~min \\
50~MB & 42.01 & \$63{,}009 & 18--73~h & \$2{,}522 & 24.6~min & \$25.09 & 16.1~min \\
\bottomrule
\end{tabular}
\end{table}

The increment shadows the floor. Its rate exceeds the upload's 714 gas/byte by about 12\%,
so at every size a disputed Identity claim costs about 1.12$\times$ what publishing the
artifact cost, and its duration is the floor's to within four transactions, since the sweep
replays the same chunk sequence the upload wrote.

\subsubsection{OVC-Conformance}

A conformance dispute must first place the plaintext on-chain: the Consumer deposits $G$
(714 gas/byte) and adjudication then scans it for motion commands (1{,}036 gas/byte),
a combined 1.75~k gas/byte, about 2.2$\times$ the Identity increment, with the deposit
and the scan contributing in roughly equal parts on synthetic input
(Table~\ref{tab:dispute_conformance}). On the real artifact the split collapses: the scan
short-circuits on the first G0/G1 command, which unmodified slicer output supplies at the
top of every chunk, so the measured scan costs 81 gas/byte instead of 1{,}036 and the
3DBenchy increment is 55\% below the synthetic projection, dominated almost entirely by the
deposit (4.79 of 5.35~Ggas). The synthetic rows, generated adversarially with the motion
command at the end of the chunk, are the predicate's true upper bound.

\begin{table}[H]
\centering
\caption{R4 dispute increment for OVC-Conformance: one deposit of $G$ followed by one scan
sweep.}
\label{tab:dispute_conformance}
\setlength{\tabcolsep}{3pt}
\footnotesize
\begin{tabular}{lrrrrrrr}
\toprule
 & & \multicolumn{2}{c}{\textbf{Ethereum}} & \multicolumn{2}{c}{\textbf{Arbitrum}} & \multicolumn{2}{c}{\textbf{opBNB}} \\
\cmidrule(lr){3-4} \cmidrule(lr){5-6} \cmidrule(lr){7-8}
\textbf{Artifact} & \textbf{Gas (G)} & \textbf{Cost} & \textbf{Time} & \textbf{Cost} & \textbf{Time} & \textbf{Cost} & \textbf{Time} \\
\midrule
64~KB & 0.11 & \$172.08 & 4.2--16.8~min & \$6.89 & 5~s & \$0.0685 & 9~s \\
256~KB & 0.46 & \$688.31 & 12.6--50.4~min & \$27.55 & 16~s & \$0.2739 & 18~s \\
1~MB & 1.84 & \$2{,}753 & 0.8--3.0~h & \$110.18 & 1.0~min & \$1.10 & 45~s \\
5~MB & 9.18 & \$13{,}766 & 3.7--14.7~h & \$550.91 & 5.0~min & \$5.48 & 3.3~min \\
\textbf{3DBenchy} & \textbf{5.35} & \textbf{\$8{,}027} & \textbf{4.7--18.9~h} & \textbf{\$323.73} & \textbf{6.3~min} & \textbf{\$2.93} & \textbf{4.2~min} \\
10~MB & 18.35 & \$27{,}532 & 7.3--29.4~h & \$1{,}102 & 9.9~min & \$10.96 & 6.5~min \\
25~MB & 45.89 & \$68{,}831 & 18--73~h & \$2{,}755 & 24.6~min & \$27.39 & 16.1~min \\
50~MB & 91.77 & \$137{,}662 & 37--146~h & \$5{,}509 & 49.3~min & \$54.78 & 32.1~min \\
\bottomrule
\end{tabular}
\end{table}

\subsubsection{OVC-Feature}

The feature dispute is the heaviest: the same deposit of $G$, followed by evaluation of the
extruder machine model of Equation~\ref{eq:material_sum} at 2{,}231 gas/byte, a combined
2.94~k gas/byte, 3.7$\times$ the Identity increment
(Table~\ref{tab:dispute_feature}). Interpreting the witness as a machine program rather
than scanning it is what the premium buys: tokenizing every line and maintaining mode and
datum state across the four properties of production slicer output enumerated in
Section~\ref{subsubsec:baseline_impl}. The 3DBenchy increment runs 11\% below the synthetic
projection because real G-code carries fewer E-coordinates per byte than the adversarial
synthetic.

\begin{table}[H]
\centering
\caption{R4 dispute increment for OVC-Feature: one deposit of $G$ followed by one
machine-model evaluation sweep.}
\label{tab:dispute_feature}
\setlength{\tabcolsep}{3pt}
\footnotesize
\begin{tabular}{lrrrrrrr}
\toprule
 & & \multicolumn{2}{c}{\textbf{Ethereum}} & \multicolumn{2}{c}{\textbf{Arbitrum}} & \multicolumn{2}{c}{\textbf{opBNB}} \\
\cmidrule(lr){3-4} \cmidrule(lr){5-6} \cmidrule(lr){7-8}
\textbf{Artifact} & \textbf{Gas (G)} & \textbf{Cost} & \textbf{Time} & \textbf{Cost} & \textbf{Time} & \textbf{Cost} & \textbf{Time} \\
\midrule
64~KB & 0.19 & \$289.49 & 4.2--16.8~min & \$11.58 & 5~s & \$0.1154 & 9~s \\
256~KB & 0.77 & \$1{,}158 & 12.6--50.4~min & \$46.33 & 16~s & \$0.4615 & 18~s \\
1~MB & 3.09 & \$4{,}632 & 0.8--3.0~h & \$185.34 & 1.0~min & \$1.85 & 45~s \\
5~MB & 15.44 & \$23{,}159 & 3.7--14.7~h & \$926.68 & 5.0~min & \$9.23 & 3.3~min \\
\textbf{3DBenchy} & \textbf{17.55} & \textbf{\$26{,}326} & \textbf{4.7--18.9~h} & \textbf{\$1{,}052} & \textbf{6.3~min} & \textbf{\$10.68} & \textbf{4.2~min} \\
10~MB & 30.88 & \$46{,}319 & 7.3--29.4~h & \$1{,}853 & 9.9~min & \$18.46 & 6.5~min \\
25~MB & 77.20 & \$115{,}797 & 18--73~h & \$4{,}633 & 24.6~min & \$46.15 & 16.1~min \\
50~MB & 154.40 & \$231{,}594 & 37--146~h & \$9{,}267 & 49.3~min & \$92.31 & 32.1~min \\
\bottomrule
\end{tabular}
\end{table}

Two kinds of result must be separated here. The first is structural and holds whatever the
predicate: per-byte cost is flat once segmented, and the limiting resource is transaction
count rather than a single-transaction gas spike. The second is specific to the feature
function instantiated here. That the semantic predicate outweighs the cryptographic work of
OVC-Identity inverts the intuition that confidentiality machinery dominates cost, and it is
the substantive scaling result of this section; but it is a result about \emph{this} $f$. A
cheaper feature, such as a value lifted from a file header, would not bind chunk selection
at all, while a richer one such as 5-axis collision detection would bind it harder. What
generalizes is a conditional rather than a ranking: a semantically rich predicate \emph{can}
outweigh the confidentiality machinery, and where it does, it alone sets the operating
point. The predicate ordering, the per-byte constants, the operating chunks of
Table~\ref{tab:chunkselect} and everything derived from them are properties of the
implementation measured here, not of the framework.

\subsubsection{The Fully Contested Dispute}
\label{subsec:economics_results}

Table~\ref{tab:dispute_all} reports the worst case: all four instances contested at once.
The increment is the sum of the four above: two deposits of $G$, three adjudication
sweeps and the $O(1)$ Access dispute, 5.50~k gas/byte in all, or 7.7$\times$ the cost of
publishing the artifact in the first place. Measured on the real artifact, the fully contested dispute costs 28.3~Ggas against the 37.0 the worst-case synthetic rates predict for its size, 24\% less, the Conformance short-circuit and the Feature E-density effect compounding.

\begin{table}[H]
\centering
\caption{R4 dispute increment with all four instances contested: two deposits of $G$, three
adjudication sweeps, and the $O(1)$ OVC-Access dispute.}
\label{tab:dispute_all}
\setlength{\tabcolsep}{2.7pt}
\footnotesize
\begin{tabular}{lrrrrrrr}
\toprule
 & & \multicolumn{2}{c}{\textbf{Ethereum}} & \multicolumn{2}{c}{\textbf{Arbitrum}} & \multicolumn{2}{c}{\textbf{opBNB}} \\
\cmidrule(lr){3-4} \cmidrule(lr){5-6} \cmidrule(lr){7-8}
\textbf{Artifact} & \textbf{Gas (G)} & \textbf{Cost} & \textbf{Time} & \textbf{Cost} & \textbf{Time} & \textbf{Cost} & \textbf{Time} \\
\midrule
64~KB & 0.36 & \$540.43 & 12.3--49.2~min & \$21.63 & 15~s & \$0.2153 & 32~s \\
256~KB & 1.44 & \$2{,}161 & 0.6--2.2~h & \$86.49 & 44~s & \$0.8610 & 54~s \\
1~MB & 5.76 & \$8{,}645 & 1.9--7.7~h & \$345.96 & 2.6~min & \$3.44 & 2.0~min \\
5~MB & 28.82 & \$43{,}227 & 9.2--36.9~h & \$1{,}730 & 12.4~min & \$17.22 & 8.4~min \\
\textbf{3DBenchy} & \textbf{28.30} & \textbf{\$42{,}451} & \textbf{11.8--47.3~h} & \textbf{\$1{,}699} & \textbf{15.9~min} & \textbf{\$16.86} & \textbf{10.8~min} \\
10~MB & 57.64 & \$86{,}453 & 18--74~h & \$3{,}460 & 24.7~min & \$34.44 & 16.4~min \\
25~MB & 144.09 & \$216{,}133 & 46--183~h & \$8{,}649 & 1.0~h & \$86.09 & 40.4~min \\
50~MB & 288.18 & \$432{,}266 & 92--366~h & \$17{,}298 & 2.1~h & \$172.18 & 1.3~h \\
\bottomrule
\end{tabular}
\end{table}

Table~\ref{tab:instance_cost} decomposes the 3DBenchy increment by instance.
\textbf{OVC-Feature alone accounts for 62\% of a fully contested dispute}, more than the
other three combined, against 19\% for OVC-Identity, even though the latter performs the
byte-wise decryption of the entire ciphertext. This is the economic expression of the
scaling result above: on a real artifact the semantic predicate, not the confidentiality
machinery, is the dominant cost center. Gas is reported for the Ethereum configuration;
per-chain totals differ by less than 0.8\% because total gas is nearly chunk-independent,
whereas cost differs by three orders of magnitude because gas price is not.

\begin{table}[H]
\centering
\caption{Per-instance decomposition of the fully contested 3DBenchy dispute increment
(6.41~MB).}
\label{tab:instance_cost}
\setlength{\tabcolsep}{3.5pt}
\footnotesize
\begin{tabular}{lrrrrr}
\toprule
\textbf{Instance} & \textbf{Gas (G)} & \textbf{Share} & \textbf{Ethereum} & \textbf{Arbitrum} & \textbf{opBNB} \\
\midrule
OVC-Access      & 0.0001 & $<$0.1\% & \$0.10 & \$0.0038 & \$0.00004 \\
OVC-Identity    & 5.398  & 19.1\% & \$8{,}098 & \$323.91 & \$3.24 \\
OVC-Conformance & 5.351  & 18.9\% & \$8{,}027 & \$321.09 & \$3.21 \\
OVC-Feature     & 17.551 & 62.0\% & \$26{,}326 & \$1{,}053 & \$10.53 \\
\midrule
\textbf{Total}  & \textbf{28.30} & \textbf{100\%} & \textbf{\$42{,}451} & \textbf{\$1{,}698} & \textbf{\$16.98} \\
\bottomrule
\end{tabular}
\end{table}

\begin{figure}[H]
    \centering
    \includegraphics[width=0.92\textwidth]{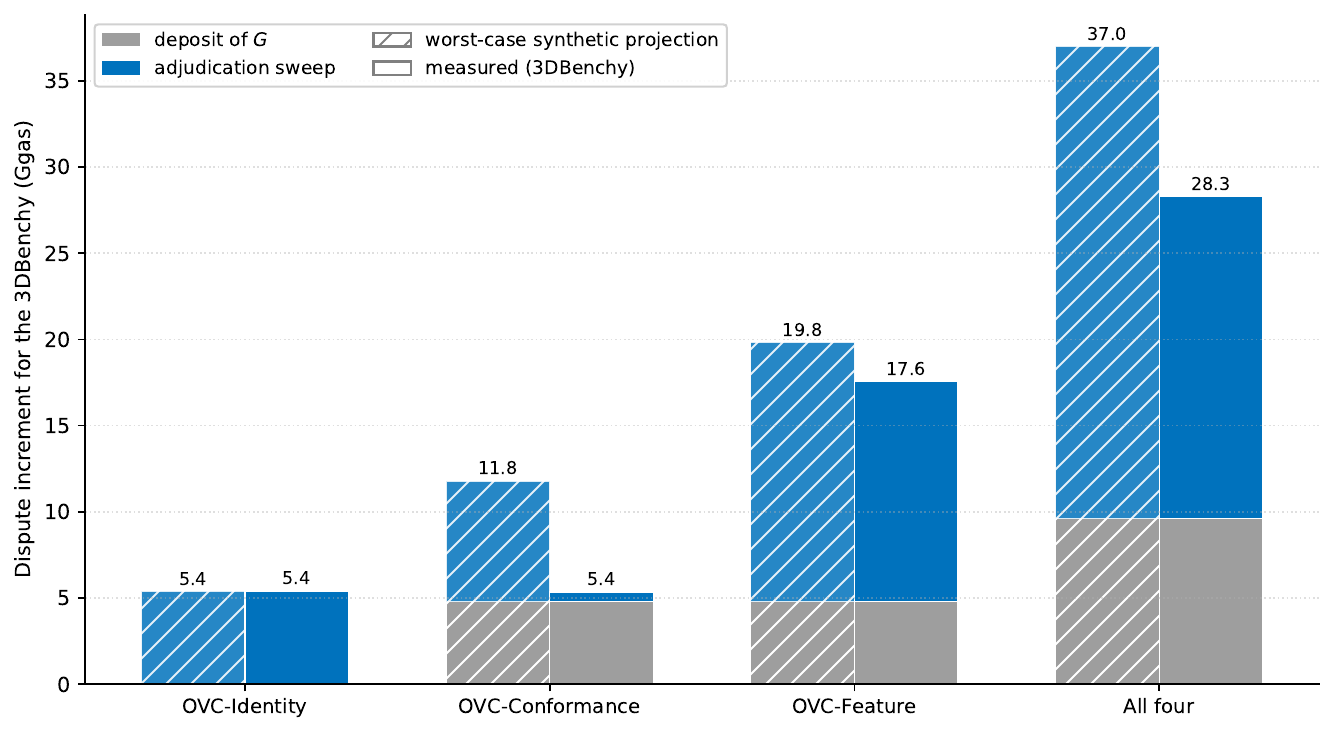}
    \caption{Composition of the dispute increment for the 3DBenchy (6.41~MB) at Ethereum's
    $c^{*}$, in gas, chain-independent up to the sub-percent chunk effect. Each instance
    is a deposit of $G$ plus an adjudication sweep; the hatched bar is the projection from
    the worst-case synthetic witnesses of Section~\ref{subsec:setup}, the solid bar the
    measurement on the real artifact. OVC-Identity needs no deposit and does not move
    between the two. OVC-Conformance's measured sweep collapses to 81 gas/byte (the scan
    short-circuits on the first motion command), leaving its dispute almost entirely
    deposit. OVC-Feature's evaluation shrinks 11\%. The all-four bar includes the $O(1)$
    OVC-Access dispute, invisible at this scale.}
    \label{fig:dispute_composition}
\end{figure}

Figure~\ref{fig:dispute_composition} draws the same structure across the instances, each
dispute split into its deposit and its adjudication sweep, the worst-case bound beside the
measurement.

\subsubsection{The Full Cost}
\label{subsec:full_cost}

Table~\ref{tab:full_cost} closes the evaluation by summing the three stages: the demand
publication of Table~\ref{tab:floor}, the binding handshake of Table~\ref{tab:handshake},
and the fully contested dispute of Table~\ref{tab:dispute_all}. Each entry is a job in
which everything goes wrong (posted, bound, and challenged on all four claims), and
the duration runs from the first posting transaction to the verdict. The undisputed
counterpart needs no table of its own: it is Table~\ref{tab:floor} plus \$1.51 or less,
on every chain, at every size.

\begin{table}[H]
\centering
\caption{R4 full protocol cost: demand publication + binding + fully contested dispute,
across the artifact ladder at the operating chunks of Table~\ref{tab:chunkselect}. The
3DBenchy row is the real 6.41~MB artifact, using per-chunk gas measured on its G-code. Gas
is quoted at Ethereum's $c^{*}$; the other chains differ by less than 0.7\%. Durations run
from posting to verdict and are ceilings in the sense of Section~\ref{subsec:metrics};
Ethereum is reported as a band ($\delta \in [1,4]$), the lower-variance rollups at
$\delta \approx 1$.}
\label{tab:full_cost}
\setlength{\tabcolsep}{2.7pt}
\footnotesize
\begin{tabular}{lrrrrrrr}
\toprule
 & & \multicolumn{2}{c}{\textbf{Ethereum}} & \multicolumn{2}{c}{\textbf{Arbitrum}} & \multicolumn{2}{c}{\textbf{opBNB}} \\
\cmidrule(lr){3-4} \cmidrule(lr){5-6} \cmidrule(lr){7-8}
\textbf{Artifact} & \textbf{Gas (G)} & \textbf{Cost} & \textbf{Time} & \textbf{Cost} & \textbf{Time} & \textbf{Cost} & \textbf{Time} \\
\midrule
64~KB & 0.41 & \$612.38 & 0.2--1.0~h & \$24.51 & 19~s & \$0.2439 & 39~s \\
256~KB & 1.63 & \$2{,}444 & 0.7--2.7~h & \$97.80 & 52~s & \$0.9735 & 1.1~min \\
1~MB & 6.51 & \$9{,}771 & 2.3--9.3~h & \$390.99 & 3.1~min & \$3.89 & 2.5~min \\
5~MB & 32.56 & \$48{,}846 & 11.1--44.3~h & \$1{,}955 & 14.9~min & \$19.46 & 10.1~min \\
\textbf{3DBenchy} & \textbf{33.11} & \textbf{\$49{,}660} & \textbf{14.2--56.8~h} & \textbf{\$1{,}988} & \textbf{19.1~min} & \textbf{\$19.73} & \textbf{12.9~min} \\
10~MB & 65.13 & \$97{,}690 & 22--88~h & \$3{,}909 & 29.7~min & \$38.91 & 19.7~min \\
25~MB & 162.81 & \$244{,}221 & 55--220~h & \$9{,}773 & 1.2~h & \$97.27 & 48.5~min \\
50~MB & 325.63 & \$488{,}440 & 110--439~h & \$19{,}546 & 2.5~h & \$194.54 & 1.6~h \\
\bottomrule
\end{tabular}
\end{table}

A fully contested job costs 8.7$\times$ an undisputed one at the synthetic rates, and
6.9$\times$ for the 3DBenchy, whose measured dispute runs cheap. The separation between
the chains is set almost entirely by gas price rather than by protocol-level gas (approximately 25$\times$ from Ethereum to Arbitrum and 100$\times$ from Arbitrum to opBNB), so the absolute figures scale linearly with token price while the chain ordering is
invariant to it. Duration separates the chains just as sharply: the contested 3DBenchy
resolves in 12.9~minutes on opBNB and 19.1~minutes on Arbitrum, against 14.2--56.8~hours
on Ethereum, and at 50~MB the Ethereum band reaches 4.5--18~days, beyond any practical
manufacturing SLA on Layer~1. Parallelizing the four disputes across accounts would cut
the dispute portion to roughly 40\%, the wall-clock then being gated by the
deposit-then-adjudication sequence of OVC-Conformance or OVC-Feature.

\section{Discussion}
\label{sec:discussion}

The principal empirical finding is a boundary: beyond the single-transaction ceiling, monolithic verification is structurally infeasible on any real chain regardless of fee level, whereas segmented execution remains tractable to at least 50~MB --- practically on rollups, expensively on Layer~1. This section elaborates that boundary, positions the framework's confidentiality and deterrence model against alternatives, establishes the integrity of the four-instance composition, and states the system's limitations.

\subsection{The Computational Boundary and the Layer-2 Imperative}
\label{subsec:disc_scalability}

The measurements separate three costs that discussions of on-chain verification usually merge. The protocol's \emph{own} machinery is nearly free: binding all four claims and accepting them costs \$1.51 on Ethereum, six cents on Arbitrum, and a fraction of a cent on opBNB, independent of artifact size (Section~\ref{subsec:handshake_results}). What costs money is moving the artifact, and it does so twice over: \textbf{publishing} it is $O(N)$ and \emph{unconditional}, incurred before any Provider bids and whether or not any ever does (Section~\ref{subsec:floor_results}), while \textbf{re-executing} it under dispute is $O(N)$ again but conditional --- paid only if a claim is actually challenged. The asymmetry that matters for adoption is therefore not between honest and dishonest parties but between the two $O(N)$ data movements and the $O(1)$ verification logic between them: the commitment apparatus that makes an artifact adjudicable is cheap, while publishing it so that it \emph{can} be adjudicated is not. Because the Demand is provider-independent, the publication cost is paid once per artifact rather than once per Provider, and the same upload survives a re-binding if a first deal lapses.

The dispute path, by contrast, exposes a hard \textbf{computational boundary} imposed by the EVM's single-transaction gas model. The decisive observation in Table~\ref{tab:mono} is that infeasibility is driven by the large \emph{linear} term, not the superlinear tail: even the smallest industrial reveal exceeds any real block limit, and a 36~M gas budget implies a single-transaction ceiling of roughly 26~KB below every measured artifact size, with quadratic memory expansion only worsening an already infeasible position. Segmented adjudication crosses this boundary by resetting EVM memory at each chunk and bounding every transaction beneath the block ceiling; it does not reduce the underlying work. Total dispute gas takes the form $G_{seg}(n,s)=O(1)+\lceil n/s \rceil\,g(s)$, linear in $n$ once the chunk size $s$ is fixed. Artifact size therefore does not break OVC; the only decisive question is whether one chunk fits inside the target chain's block gas limit.

Segmentation is not free (each extra chunk adds transaction overhead), but the best chunk is the \emph{largest} feasible one, not the smallest: across the feasible range total gas varies by about 2\% while transaction count varies twentyfold, making chunk selection a chain-parameterized systems problem rather than an algorithmic one. The consequence compounds across cost and availability: at 50~MB a fully contested dispute adds \$432{,}266 and roughly four days at best on Ethereum, against \$172 and about 1.3~hours on opBNB (Table~\ref{tab:full_cost}) --- the difference between a multi-day service freeze and a bounded operational delay.

Layer-2 deployment is therefore a \textbf{liveness requirement}, not merely a cost optimization. Without rollup-scale block limits, segmentation would only replace one impossible monolithic transaction with thousands of feasible but operationally unmanageable ones; higher-throughput layers admit larger chunks (only opBNB's 200~M block reaches 80~KB), collapse the transaction count, and lower the penalty of each sequential step. The EIP-1559 fee mechanism~\cite{buterin_eip1559_2019} compounds the L1 disadvantage: to avoid multi-block exclusion a participant must raise the priority tip, directly inflating $C_{real}$, whereas a rollup's lower base fee leaves margin to keep the delay factor $\delta$ near unity. The result is a clean architectural separation: L1 for settlement anchoring, L2 as the only practical execution surface for multi-transaction verification.

\subsection{Confidentiality, Deterrence, and Positioning}

OVC provides \emph{optimistic confidentiality}: the committed artifact stays undisclosed on the no-dispute path. A key subtlety governs what disclosure under dispute actually exposes. Pricing itself never requires the artifact: a Provider bids from the demand tuple alone, quoting on the declared feature value $f_{claim}$ without seeing the design. The plaintext becomes readable only to the \emph{selected} Provider, who alone can decrypt the key published at Binding and recover $G$ from the ciphertext, so that party can verify it before committing machine time. A dispute is therefore raised by a Provider who already holds the plaintext, so it discloses nothing new \emph{to that Provider}; what it forces is an \emph{on-chain, public} reveal, exposing the IP to the wider market irrevocably. The asset OVC protects is thus market-wide secrecy, and disclosure under dispute is the deliberate price of proving a contested claim. A Consumer who values that secrecy above proving the claim may decline to disclose, resolving to \textsc{Withheld} (a terminal state distinct from the fraud outcomes): this forgoes the contract but denies any public leak. OVC is therefore a \textbf{verification primitive}, not a complete deterrence system: it produces a deterministic verdict, and the deterrent effect is supplied by the surrounding marketplace.

This mirrors the coexistence of optimistic and zero-knowledge rollups. Zero-knowledge adjudication offers stronger confidentiality (it can prove a predicate without revealing the witness) but generally incurs proving cost on every claim, including honest ones, whereas OVC charges no proving cost on honest claims and incurs verification cost only under dispute, at the price of publishing the artifact in encrypted form up front. The two are complementary operating points rather than competitors, and a ZK adjudication path is a natural extension we leave to future work (Section~\ref{sec:future_work}).

Realizing OVC on a blockchain is justified not because the predicate logic is impossible elsewhere, but because the \emph{outcome} must be both produced without a trusted operator and consumed by a separate economic layer. A public append-only log can anchor commitments but cannot execute the predicate; a trusted execution environment can execute it but shifts trust to hardware vendors and attestation; traditional arbitration can decide but is jurisdiction-bound, slow, and not machine-consumable. Blockchain uniquely combines immutable commitment, neutral deterministic execution, and a public verdict that an external economic or social layer can consume without trusting any operator. That consumption is where deterrence lives: an adverse verdict (\textsc{Incorrect} or \textsc{Withheld}) can carry an automatic economic consequence, while the permanent on-chain record lets a market avoid actors with a history of such verdicts. \textbf{IP-probing} --- a dispute raised solely to force a public reveal --- is \emph{bounded} rather than prevented. The protocol cannot detect a probe: a challenge carries no statement of motive, and an insincere one is indistinguishable from a sincere one. What it does guarantee is that the probe extracts nothing, because disclosure is consent-gated. Publishing the witness on-chain requires the Consumer's own deposit and adjudication transactions, so no Provider can compel a reveal; a Consumer facing a probe may simply decline, and the claim times out to \textsc{Withheld}. The guarantee holds whenever the probe is raised, before or after the artifact became readable to the selected Provider; it is the consent gate, not prior possession, that carries the defense. The residual exposure is therefore griefing rather than disclosure: withholding forfeits the contract, so a probe can deny a deal it still cannot read, and the \textsc{Valid} verdict records the challenge against its initiator. The framework supplies the deterministic verdict and its attribution; calibrating the economic and social consequences is the responsibility of the surrounding marketplace.

\subsection{Pipeline Integrity}

Decomposing negotiation into four instances introduces no new substitution attack, and splitting the lifecycle into bidding and binding does not open one. The shared commitments are coordinator-enforced at bind time (Section~\ref{subsubsec:binding_phase}): the key $K$ is pinned by $H(K)$ across OVC-Access and OVC-Identity, the artifact $G$ by $\widetilde{H}(G)$ across Identity, Conformance, and Feature, and the delivery triple by $c_{ID}=H(K\|G\|C)$. The binding transaction does not accept the artifact values from its caller: it reads them from the sealed Demand, so the instances are necessarily bound to the artifact that was published for bidding. Passing two stages with different values of a shared parameter would require a hash collision, so the feature extracted in the final stage is provably derived from the same bitstream verified in the earlier stages, and that bitstream is the one Providers priced against. The pipeline's security thus reduces to the security of each instance plus the one-time, publicly checkable correctness of the \texttt{BindingPhase} coordinator, ruling out ``modular bait-and-switch'' attacks under the collision-resistance assumption.

\subsection{Limitations and System Boundaries}
\label{subsec:limitations}

While the OVC establishes a sound mathematical primitive for verification, its scope is intentionally bounded to the veracity of digital claims. Several factors remain external to the protocol and must be addressed by the surrounding marketplace architecture.

First, the protocol is strictly limited to \textbf{deterministically verifiable claims}. The OVC operates on the digital representation of the manufacturing task (the G-code); it cannot, by itself, verify physical properties that emerge during the execution phase, such as final surface roughness or material porosity. These post-production qualities require alternative dispute mechanisms or the integration of real-time sensor oracles.

Second, there is a structural trade-off between \textbf{predicate complexity and computational cost}. Evaluating declared material consumption on unmodified slicer output requires a stateful machine model rather than a syntactic scan, which alone makes OVC-Feature the heaviest instance and the binding constraint on chunk size for every chain considered (Table~\ref{tab:chunkselect}); richer predicates such as 5-axis collision detection would compound the effect. The bottleneck is then not the block gas limit but adjudication gas itself: if the cost of adjudication exceeds the economic value of the contract, the protocol is impractical for low-margin tasks. The 3DBenchy measurement (\$16.86 on opBNB with all four instances disputed) suggests the current predicates remain viable for professional print contracts on rollups, while Ethereum excludes low-margin work. The results are specific to the implemented predicates, and the real-artifact margins below the worst-case synthetic bounds are measured on a single canonical artifact: the bounds hold for any input by construction, but the margin's size --- 24\% for the fully contested 3DBenchy dispute --- varies with geometry, slicer, and print profile.

Third, the cost asymmetry between depositing an artifact and adjudicating it admits a
\textbf{configuration hazard} that nothing in the contracts prevents. The deposit is roughly
three times cheaper per byte than the heaviest predicate (Table~\ref{tab:chunkbound}), so a
chunk may be comfortably uploadable and yet unadjudicable: on Ethereum any choice between
roughly 15 and 47~KB deposits without difficulty and then strands the dispute, since
\texttt{processChunk} can never be included in a block. The Consumer gains nothing by this (the claim times out to \textsc{Withheld} and the contract is forfeited), but an honest
Consumer must choose correctly. The chunk size is publicly observable before any Provider
bids, so a Provider can screen a Demand for adjudicability on its target chain, which places
the defense in the marketplace layer alongside the framework's other economic consequences.

Fourth, posting the encrypted artifact at bidding makes \textbf{artifact size} public. The ciphertext is byte-for-byte as long as the G-code, so any observer learns the size of the concealed job, which correlates loosely with print duration and part complexity. An adjudicable delivery channel requires the ciphertext to be observable, so the leak follows from putting delivery on the common ledger. It is coarse --- size alone, not geometry --- and is the price of removing the off-chain channel; padding the ciphertext to fixed size buckets would close it at a proportional increase in the dominant honest-path cost, a trade-off we leave to future work.

Fifth, OVC offers no \textbf{post-handover secrecy}: once the Specification Handover completes, the selected Provider holds the plaintext artifact off-chain, and its subsequent protection is a contractual and legal matter outside the protocol. Relatedly, the commitment scheme is binding but not hiding (Section~\ref{subsec:atomic_ovc}), and which instances this exposes follows from what each one commits to. OVC-Access and OVC-Identity embed the secret key $K$ in their witness, directly or through $H(K \| G \| C)$, so their commitments are effectively blinded against a guesser who does not hold $K$. OVC-Conformance and OVC-Feature commit to the plaintext artifact itself, so their unsalted plaintext-hash commitments admit dictionary-style guess confirmation for low-entropy artifact families. The gap is therefore confined to the two artifact predicates. For multi-kilobyte G-code the witness space is large in practice, but designers extending OVC to low-entropy witnesses should treat this as a real confidentiality boundary rather than a theoretical caveat; Section~\ref{sec:future_work} gives the remedy.

Finally, OVC verifies \textbf{claim veracity}, not \textbf{service reliability}: it ensures the Provider receives the correct file and that the Consumer's claims about it are true, but it does not guarantee that the Provider executes the manufacturing service competently or honestly. Bidder selection and execution assurance remain the province of the external economic and social subsystems discussed in Section~\ref{sec:discussion}, which consume the OVC verdict as evidence.

\section{Future Work}
\label{sec:future_work}

Six directions follow naturally from the framework's boundaries.

\textbf{Zero-Knowledge Adjudication:}
The strongest confidentiality limitation (disclosure of the witness under dispute) is removed by zero-knowledge adjudication, in which a claimant proves $\mathcal{P}(w, c)$ via succinct cryptographic arguments rather than revealing $w$. The predicates formalized in this work provide a direct specification for such circuits, and the Specification Handover abstraction is mechanism-neutral: a ZK adjudication path preserves the phase structure and lifecycle semantics of Table~\ref{tab:lifecycle}. The trade is the always-on proving cost discussed in Section~\ref{sec:discussion}.

\textbf{Hardened Commitments and Selective Disclosure:}
Salting the plaintext-hash commitments of OVC-Conformance and OVC-Feature ($H(G \| r)$) would close the dictionary-confirmation gap identified in Section~\ref{subsec:limitations}, and selective disclosure revealing only the witness regions needed to settle a dispute would reduce, though not eliminate, disclosure for scan-type predicates.

\textbf{Interactive Fault Attribution and Richer Predicates:}
Multi-round interactive challenge, in which the challenger localizes a fault to a specific index or region, would reduce adjudication from $O(n)$ to $O(1)$ in many cases and sharply lower the cost of large-artifact disputes. This in turn makes a richer predicate library affordable (geometric constraints, bounding-volume or tolerance checks, toolpath-derived time estimates), extending OVC beyond the four demonstrator predicates without the per-byte scan cost that currently bounds predicate complexity.

\textbf{Custom and Hybrid Execution Layers:}
The results identify the execution layer, not the protocol, as the limiting constraint on large-artifact disputes (Section~\ref{subsec:disc_scalability}), motivating co-design of the execution surface rather than deployment onto a general-purpose chain unchanged. Native precompiles for the dispute kernels would reduce the dominant inner loops rather than relocating them across transactions as segmentation does; a dedicated data-availability path for the witness (blob transactions or an external DA layer) attacks the per-byte calldata term while the settlement chain retains only the verdict and the rolling-hash anchor; and parallel chunk execution generalizes the cross-instance parallelism of Section~\ref{subsec:full_cost} to within-instance chunks. These reduce the cost of verifying a given amount, where the algorithmic directions above reduce what must be verified. The open tension is that bespoke appchains and TEE-assisted hybrids buy throughput by narrowing the validator set or shifting trust to hardware, eroding the neutral, operator-free execution that makes an OVC verdict consumable without trust; the research question is how to recover the throughput without surrendering that neutrality.

\textbf{Empirical Cost Characterization on a Real Corpus:}
The evaluation bounds cost with worst-case synthetic witnesses and samples the real-artifact margin at one canonical model. A corpus study --- real G-code spanning geometry classes (calibration, mechanical, organic, vase-mode), multiple slicers, and print profiles --- would turn that margin into a distribution: how far below the bound typical jobs fall per predicate, and which artifact properties drive the variation. Because the measurement harness already accepts arbitrary G-code, the study is mechanical; its value is the external validity of the economic thresholds, which this paper states as bounds rather than expectations.

\textbf{Incentive Modeling:}
A formal analysis of the economic and social consequences attached to verdicts, and of the resulting dispute incentives, would allow quantitative evaluation of equilibrium behavior under rational adversaries and strengthen the protocol's deterrence guarantees.

\section{Conclusion}
\label{sec:conclusion}

This paper introduced the \textbf{Optimistic Verifiable Claim (OVC)} framework, a modular verification primitive for decentralized manufacturing negotiation. OVC is not a marketplace or a deterrence system but a deterministic adjudication mechanism: it binds a claim about a concealed artifact at bidding time and renders it provably falsifiable under dispute, producing a cryptographically grounded verdict while keeping the artifact confidential on the honest path. In practice this lets a Provider trust a published demand (say, that a concealed design consumes a stated length of filament) enough to price it and return a binding offer, as though the specification were disclosed, yet with any falsehood provably exposed upon contestation and witness disclosure. Bidding proceeds on the claim alone: the design never changes hands at bid time, and no trusted intermediary need vouch for it.

Decomposing negotiation into a pipeline of Access, Identity, Conformance, and Feature claims enables claim-specific adjudication beyond content-agnostic fair exchange. The evaluation separates three costs that are usually conflated: verification itself is nearly free and independent of artifact size; publishing the artifact so that delivery is adjudicable is $O(N)$ and unconditional; re-executing it under dispute is $O(N)$ and conditional. Monolithic verification meets a concrete \textbf{computational boundary} --- it exceeds any real block gas limit beyond a few tens of kilobytes --- and segmented adjudication relocates the work across bounded transactions, so artifact size is limited by the execution layer rather than by the protocol, with rollups the only practical surface for industrial-scale disputes. On the 6.41~MB 3DBenchy this is the difference between publishing for \$2.87 and disputing for \$16.86, both within minutes, on a high-throughput rollup, against \$7{,}207 and \$42{,}451 over up to 57~hours in total on Ethereum. A second finding is methodological: verifying a declared physical quantity against real G-code requires interpreting it as a machine program, and that semantic work exceeds the cryptographic work of decryption and hashing. Where a predicate is semantically rich in this way, the domain predicate --- not the confidentiality machinery --- sets the operating point for the whole pipeline; the measured ordering is a property of the feature function instantiated here, and the framework admits any deterministic, publicly specified, on-chain-evaluable feature.

OVC is deliberately bounded: it verifies claim veracity, not service reliability or correct physical execution, and its confidentiality is optimistic: a contested claim is proven by public disclosure. This work addresses the three pre-contractual phases of Table~\ref{tab:lifecycle}; execution, physical delivery, validation, and settlement lie outside the protocol and remain to be integrated, for instance through sensor oracles and on-chain settlement that consume the OVC verdict. Within those boundaries the framework contributes a primitive for optimistically confidential, claim-aware negotiation, expanding blockchain coordination from atomic delivery toward structured adjudication and supplying the deterministic verdict on which external economic and social mechanisms can act.

\section*{Code and Data Availability}
The smart contracts, the gas-measurement test suite, and the raw measurement results that support the findings of this study are openly available at \url{https://github.com/fsprojekti/optimistic-verifiable-claims} and permanently archived on Zenodo~\cite{corn_ovc_2026} (DOI:~10.5281/zenodo.20794987).

\section*{Funding}
Slovenian Research Agency: P2-0270; Ministry of Higher Education, Science and Technology of the Republic of Slovenia: 100-15-0510.

\section*{CRediT authorship contribution statement}
\textbf{Marko Corn:} Conceptualization, Methodology, Software, Validation, Formal analysis, Investigation, Visualization, Writing -- original draft. \textbf{Nejc Ro\v{z}man:} Methodology, Validation, Writing -- review \& editing. \textbf{Primo\v{z} Podr\v{z}aj:} Supervision, Resources, Funding acquisition, Project administration, Writing -- review \& editing.

\section*{Declaration of competing interest}
The authors declare that they have no known competing financial interests or personal relationships that could have appeared to influence the work reported in this paper.

\section*{Declaration of generative AI and AI-assisted technologies in the manuscript preparation process}
During the preparation of this work the authors used Claude (Anthropic) in order to improve the language, style, and readability of the manuscript. After using this tool, the authors reviewed and edited the content as needed and take full responsibility for the content of the published article. The conceptual design, methodology, implementation, results, and all scientific content are entirely the authors' own original work.

\bibliographystyle{elsarticle-num}
\bibliography{references}

\end{document}